\documentclass[11pt,draftcls,onecolumn]{IEEEtran}
\usepackage{graphicx}
\usepackage{amssymb}
\usepackage{subfigure}
\usepackage{amsfonts}
\usepackage[percent]{overpic}
\usepackage{mathrsfs,amsmath} 
\usepackage{epstopdf}
\usepackage{tabularx}
\graphicspath{{figs/}}
\usepackage{algorithm}
\usepackage{algorithmic}
\usepackage{cite}

\usepackage[para,online,flushleft]{threeparttable}

\DeclareMathOperator*{\argmin}{\arg\!\min}

\newcommand\undermat[2]{%
  \makebox[0pt][l]{$\smash{\underbrace{\phantom{%
    \begin{matrix}#2\end{matrix}}}_{\text{$#1$}}}$}#2}

\begin{document}                                    
\title{Semi-supervised Speech Enhancement in Envelop and Details Subspaces}
\author{Pengfei~Sun,
        Jianhong~Xu,
        and~Jun~Qin,} 
\maketitle

\begin{abstract}
In this study, we propose a modulation subspace (MS) based single channel speech enhancement framework, in which the spectrogram of noisy speech is decoupled as the product of a spectral envelop subspace and a spectral details subspace. This decoupling approach provides a method to specifically work on elimination of those noises that greatly affect the intelligibility. Two supervised low-rank and sparse decomposition schemes are developed in the spectral envelop subspace to obtain a robust recovery of speech components. A Bayesian formulation of non-negative factorization is used to learn the speech dictionary from the spectral envelop subspace of clean speech samples. In the spectral details subspace, a standard robust principal component analysis is implemented to extract the speech components. The validation results show that compared with four speech enhancement algorithms, including MMSE-SPP, NMF-RPCA, RPCA, and LARC, the proposed MS based algorithms achieve satisfactory performance on improving perceptual quality, and especially speech intelligibility. 
\end{abstract}
\begin{IEEEkeywords}
speech enhancement, modulation subspace, supervised, low-rank, sparsity, speech intelligibility.
\end{IEEEkeywords}
\IEEEpeerreviewmaketitle

\section{Introduction}
\IEEEPARstart{S}{peech} enhancement is a common topic in speech processing and front end of speech recognition. In general, the goal of speech enhancement is to improve both perceived quality and intelligibility by reducing residual noise while minimizing speech signal distortion. Speech with higher quality is more comfortable for audiences, and comparatively higher intelligibility is measured by lower word recognition error rates. Most of the existing speech enhancement algorithms can achieve high speech quality but relatively low performances on speech intelligibility \cite{loizou2011reasons}. 

Speech intelligibility is mainly affected by vocal tracts, which can be translated as the spectral envelop \cite{schroeder1985code}. The fact that mel-frequency cepstrum coefficients (MFCC) derived from spectral envelop is efficient in automatic speech recognition algorithms \cite{hinton2012deep} further demonstrates the importance of spectral envelop on intelligibility improvement. Generally, the human speech $X$ is a convolution acoustic procedure in the time domain, corresponding to that vocal excitation (harmonics) is modulated by vocal tract (formants) in the frequency domain: 
\begin{equation}
X = X_{\mathit{e}} \circ X_{\mathit{d}}
\label{eqxd}
\end{equation}
where $X_{\mathit{e}}$ as the 'carrier' modulates the fine-structure $X_{\mathit{d}}$, and $\circ$ is the Hadamard product. Equivalently, we can decompose the noisy speech as $Y = Y_{\mathit{e}} \circ Y_{\mathit{d}}$, where $Y_{\mathit{e}}$ is the envelop matrix and $Y_{\mathit{d}}$ is the details matrix. In the cepstrum domain, along the pseudo-frequency axis, the vocal tract $X_{\mathit{e}}$ as the identity of the speech is located at low frequency region, and the pitch $X_{\mathit{d}}$ is concentrated as high frequency components \cite{oppenheim2004frequency}. This physical modulation is naturally translated as the correlations in time-frequency (T-F) domain. Despite several works \cite{} take into account the interframe correlation, estimation of $X$ by operating on spectrogram cannot provide insightful illustration on the long term correlation, which generally is reflected by the envelop $X_{\mathit{e}}$. This may also explain that conventional spectrogram estimation sheds little light on intelligibility improvement. Therefore, decoupling the formants and pitches into different subspaces helps to process the correlations independently, and as a result may provide a better approximation to envelop matrix. 

To recover the speech spectrogram $X$, our approach is to extract speech components (i.e., $X_{\mathit{e}}$ and $X_{\mathit{d}}$) from two subspaces $Y_{\mathit{e}}$ and $Y_{\mathit{d}}$ separately. Considering that previous studies \cite{duan2012speech} show that supervised methods perform quite well in subspace speech enhancement, we propose a semi-supervised framework combining dictionary and non-dictionary based low-rank and sparse decomposition (LSD). In envelop subspace $Y_{\mathit{e}}$, driven by the motivation of improving intelligibility and the fact that speech bases may highly overlap with the convex hull of noise bases, two specifically designed algorithms, referring as two-layer LSD (TLSD-MS) and single-layer LSD (SLSD-MS), are comparatively introduced to implement the speech extraction. An offline trained speech envelop dictionary is utilized in both TLSD-MS and SLSD-MS. In $Y_{\mathit{d}}$, a general unsupervised LSD is used to obtain speech components $X_{\mathit{d}}$. The spectrogram of estimated speech can be obtained as the element-wise product of the two extracted sub matrices.

\subsection{Related Work} 
In this study, our proposed speech enhancement algorithm can be categorized as modulation subspace based semi-supervised LSD. Modulation domain based source separation technologies are mainly developed according to the knowledge that the spectrogram of speech can be described as a time-varying weighted sum of component modulations \cite{elliott2009modulation}. By exploiting intrinsic decomposition through well convex optimization, low-rank and sparse analysis overcomes the high sensitive of the conventional principle component analysis (PCA) when subjecting to large corruptions. 
 
\subsubsection{Low rank and sparse decomposition}  
The idea of applying LSD to separate speech from background noise is derived from the intrinsic data structure of noisy speech spectrogram \cite{sun2016low}, in which background noise usually demonstrates low spectral diversity whereas speeches are more instantaneous and changeable. Specific constraints (e.g., masking threshold, noise rank, and block-wise restrictions) \cite{yang2012sparse, sun2014novel} are incorporated to optimize the decomposition. LSD has been also implemented in wavelet packet transform domain\cite{robel2007cepstral, bouzid2016speech}, in which the speech components are concentrated to be more sparsity. 

In many relevant cases, using a single spectral model to describe the speech signal is insufficient. Because with long-term repeated structure, speech can also demonstrate low-rank characteristic as well as sparsity. The coexisting of low-rank and sparse properties in speech requires a more comprehensive constraint to reflect its spectral structure. Chen \cite{chen2013speech} utilized a modified robust PCA (RPCA) optimization function, in which offline trained speech spectral dictionary is employed and outlying entries are subjected to minimal energy restriction. Duan $et.~al$ introduced an online learned dictionary to implement non-negative spectrogram decomposition \cite{duan2012speech}. Yang proposed a LSD strategy via combining dictionaries with respect to speech and noise \cite{yang2013low}. Despite the supervised RPCA relying on pre-learned dictionary, to some extent, is quite similar to dictionary based non-negative matrix factorization (NMF) technique \cite{mohammadiha2013supervised} and sparse coding approach \cite{sigg2012speech}, it specifically imposes rank constraints on background noise spectra, and is more flexible and effective for non-stationary noise cancellation. 

\subsubsection{Modulation Domain based Source Separation}  
In a speech enhancement framework, the time and frequency modulations in spectrogram are intuitively represented as the correlations among neighboring spectral magnitudes. These correlations have been frequently employed as a prior knowledge to improve either the noise power estimation \cite{gerkmann2012unbiased} or speech magnitude estimation \cite{cohen2003noise}. Typically, by incorporating 1D smooth coefficients \cite{martin2001noise} or 2D average window \cite{gerkmann2008improved} imposed on the spectrogram, significant improvements on speech quality can be achieved by taking the correlations into account. 

Instead of locally introducing correlations deriving from the modulations in spectrogram, a more straightforward way is to decouple the modulation by transforming into the cepstral domain. By utilizing pseudo frequencies, Deng \(et.~al\)  \cite{deng2004estimating} conducted a conditional minimum mean square error (MMSE) estimation in the cepstrum domain, and the result showed that it was a noise-robust feature selection approach. Breithaupt \(et.~al\) \cite{breithaupt2007cepstral} proposed a higher cepstral coefficients smoothing scheme, in which the recursive temporal smoothing was only applied to the fine spectral structure. Gekmann \(et.~al\) enforced the statistical estimation in the temporal cepstral domain, and successfully obtained a more accurate speech presence probability estimation \cite{gerkmann2010speech}. Veisi and Sameti introduced hidden markov models into the mel-frequency domain \cite{veisi2013speech}, and the results indicated a significant improvement on noise cancellation. Different from cepstrum based algorithms, Paliwal \(et al.\) \cite{paliwal2010single, paliwal2012speech} proposed a frame-wise transformation along the time axis, and in the modulation domain, the clean speech is obtained based on conventional speech estimator. 

\subsection{Method Overview}
The proposed method of decomposing spectrogram into two modulation subspace has two major advantages: 1) the decoupling intrinsically incorporates the correlations existed in the speech spectrogram; 2) it strengthens the acoustic characteristics of speech components in each subspace, and makes speech components more distinctive compared with noise components. To obtain the two modulation subspaces, a cepstrum based modulation inverse (CMI) transform is applied. It firstly obtains cepstrogram by applying element-wise logarithm and discrete Fourier transform (DFT), then window functions are used to separate the envelop and details subspaces in the cepstrum domain, and finally inverse Fourier transform is implemented to obtain two modulation subspaces \cite{simsekli2014non}. 

\begin{figure}[!hbt]
\vspace{-2mm}
\includegraphics[scale=0.55]{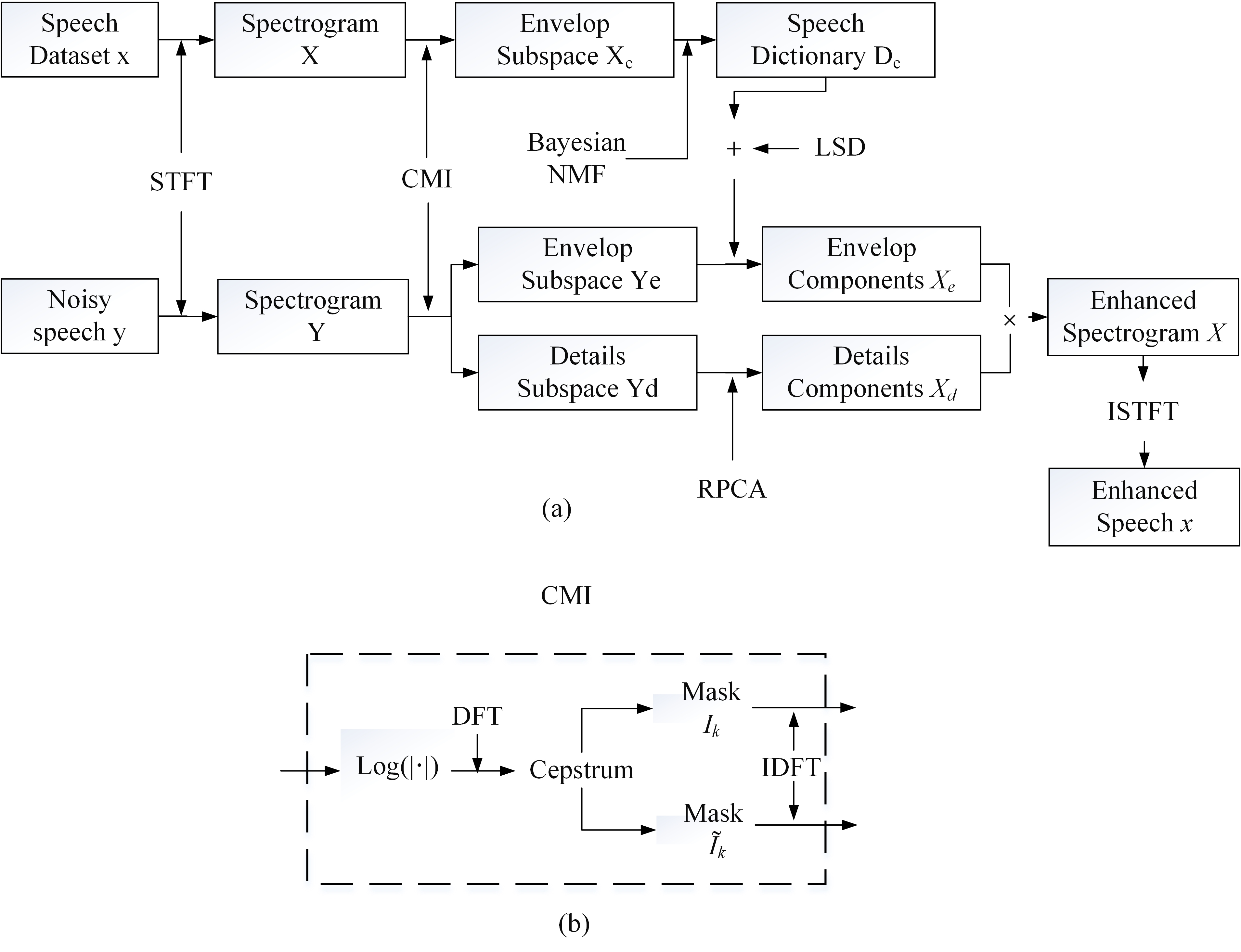}
\vspace{-4mm}
\caption{The schematic diagram of the proposed low-rank and sparse decomposition in modulation subspace (LSD-MS) algorithm (a) and details of the proposed CMI procedure (b).}
\label{schemtic}
\end{figure}

In each modulation subspace, LSD are implemented to extract the speech components. Considering that the spectral envelop subspace has a slowly varied property, noise components in this subspace share more spectral bases with speech components than that in the spectral details subspace. Therefore, in the spectral envelop subspace, supervised LSD can be implemented, in which two different decomposition strategies adapting to different types of noises are proposed. In the spectral details subspace $Y_{d}$, the speech components show highly regular structure (i.e., fine structure), and comparatively noise is supposed to be low rank. Therefore, a typical unsupervised RPCA method can be used to effectively extract the speech spectral details. Specifically, for unvoiced segments, the supervised LSD in envelop subspace and the unsupervised LSD can in details subspace can both work efficiently to minimize the residual noises as the general LSD approaches conducted in spectrogram domain. Especially, considering that the details subspace provide a more concentrated speech structure, it will yield better results comparing with the general spectrogram decomposition. The implementation procedure is shown in Fig.~\ref{schemtic}. 

\subsection{Contribution of Our Work}
By decoupling the spectral envelop and spectral details subspaces, LSD is implemented in both subspaces in this study. The contributions of the proposed algorithms can be summarized as follows:
\begin{itemize}
\item A uniform acoustic-model based framework is proposed, which naturally inherits the correlations demonstrated in speech spectrogram. In other words, the decoupling of the spectral envelop and details subspaces can help to independently reduce the distortions in these two uncorrelated subspaces. 
\item New semi-supervised speech enhancement algorithms are proposed based on the two modulation subspaces, which provides robust features for dictionary learning. 
\item Two different LSD schemes are developed to be adaptive to different background noise in the spectral envelop subspace.  
\item The proposed algorithms provide highly efficient and robust solutions to single channel speech enhancement, and the comprehensive evaluation results demonstrate significant improvements on speech quality, and particularly with respect to intelligibility, compared with existing state-of-the-art algorithms.  
\end{itemize}

The rest of the paper is organized as follows: the modulation subspace framework is presented in Section II. Section III describes the algorithms of two proposed semi-supervised LSD. In Section IV, the experiments and results are presented with the developed approaches. Finally, Section V concludes the study. 

\section{Modulation subspace}
For a speech signal corrupted by an additive noise, i.e., $Y^{c} = X^{c} + N^{c}$, the squared magnitude spectrum is given as 
\begin{equation}
\begin{split}
|Y^{c}|^2 &  = |X^{c}|^2+|N^{c}|^2 \\
 &+ 2|X^{c}||N^{c}|cos(\theta_{\Delta})
\end{split}
\end{equation}
by using $Y = |Y^{c}|$, $X = |X^{c}|$, and $N = |N^{c}|$, it can further be presented as
\begin{equation}
\begin{split}
Y &  = (X + N) \\
 &\cdot \sqrt{1+\frac{2\gamma}{(1+\gamma)^{2}}(cos(\theta_{\Delta})-1)} \\
 & = X+ N + \alpha(\gamma,\theta_{\Delta})(X+N) = X+ N^{'}
\end{split}
\label{eqspectrum}
\end{equation}
where $\gamma = \frac{N}{X}$, $\theta_{\Delta} = \theta_{X^{c}}-\theta_{N^{c}}$, and $N^{'}$ includes the speech-noise cross-term $\alpha(\gamma,\theta_{\Delta})(X+N)$. Considering the spectral envelop and spectral details subspaces, the noisy speech model in the spectrogram domain is given as 
\begin{equation}
\begin{split}
Y & = Y_{\mathit{e}} \circ Y_{\mathit{d}} \\
  & = (X_{\mathit{e}}+N_{\mathit{e}}) \circ (X_{\mathit{d}}+N_{\mathit{d}}) \\
  & = X_{e} \circ X_{d} + X_{e}\circ N_{d} + N_{e} \circ X_{d} + N_{e}\circ N_{d}
\end{split}
\label{eqny}
\end{equation}
in which noisy speech spectrogram matrix $Y \in \mathbb{R}^{n\times m}$ is the element-wise product of the spectral envelop matrix $Y_{\mathit{e}}$ and spectral details matrix $Y_{\mathit{d}}$, same as the definition in (\ref{eqxd}). Accordingly, the noise term $N^{'}$ in (\ref{eqspectrum}) is equal to $X_{e}\circ N_{d} + N_{e} \circ X_{d} + N_{e}\circ N_{d}$. $N_{e}$ and $N_{d}$ are the relative noise components in two subspaces. Note that $N_{e}\circ N_{d}$ is not necessarily equal with $N$. To conduct the decomposition in each subspace and extract $X_{\mathit{e}}$ and $X_{\mathit{d}}$, both subspaces $Y_{\mathit{e}}$ and $Y_{\mathit{d}}$ can be obtained based on the noisy speech spectrogram $Y$. For clean speech spectrogram $X$, applying window functions in the cepstral domain can effectively obtain the spectral envelop and details subspaces. However, noise greatly affects the boundaries of two subspaces in the cepstral domain \cite{openshaw1994limitations}. 

\subsection{Two Modulation Subspaces Decomposition}
In this study, the proposed CMI is applied to obtain the two modulation subspaces. Hilbert transform as a typical approach for demodulation is also used as a comparison. Unlike the Hilbert transform, CMI requires no assumption on speech details structure. In CMI method, the spectral envelop and details matrices of the noisy speech can be written as:
\begin{equation}
Y_{e} = exp(W^{-1}(H_{k}\circ (W log(Y))))
\label{eq4}
\end{equation}
\begin{equation}
Y_{d} = exp(W^{-1}(\tilde{H}_{k}\circ (W log(Y))))
\label{eq5}
\end{equation} 
in which $exp(\cdot)$ and $log(\cdot)$ are defined as element-wise exponential and logarithm, and $W$ refers to the DFT matrix
\[
W =
\begin{bmatrix}
W_{n}^{0}  & W_{n}^{0}  &\dots &W_{n}^{0} \\
W_{n}^{0}  & W_{n}^{1}  &\dots &W_{n}^{n-1}\\
\dots      & \ddots  &\ddots &\dots\\
W_{n}^{0}  & W_{n}^{n-2}  &\dots &W_{n}^{(n-1)(n-2)}\\
W_{n}^{0}  & W_{n}^{n-1}  &\dots &W_{n}^{(n-1)(n-1)}
\end{bmatrix} \]
where $W_{n} = e^{-2\pi i/n}$, and $n$ is the number of input signal data points. $H_{k}$ and $\tilde{H}_{k}$ are the mask matrices to select the low pseudo frequency components and high pseudo frequency components in the cepstrum domain, and defined as $H_{k} = V_{k}U^{T}$, and $\tilde{H}_{k} = \widetilde{V}_{k}U^{T}$, respectively. $U$ is a $m$-element column array with unity value. $V_{k}$, and $\widetilde{V}_{k}$ are $n$-element column array, and given as
\begin{equation}
\begin{split}
V_{k} &= [ 1 \quad \dots \quad 1 \quad 0 \quad \dots \quad 0  \quad 1 \quad \dots \quad 1] \\
\widetilde{V}_{k} &=[\undermat{k+1}{0 \quad \dots \quad 0}  \quad 1 \quad \dots \quad 1 \quad \undermat{k}{0 \quad \dots \quad 0} ] 
\end{split}
\end{equation} \\
where the index $k$ is the pseudo frequency in the cepstrum domain, and its corresponding frequency $f$ is given as $f = \frac{F_{s}}{2k}$, where $F_{s}$ is the sampling frequency. 

\begin{figure}[!hbt]
\vspace{-3mm}
\centerline{\includegraphics[scale=0.45]{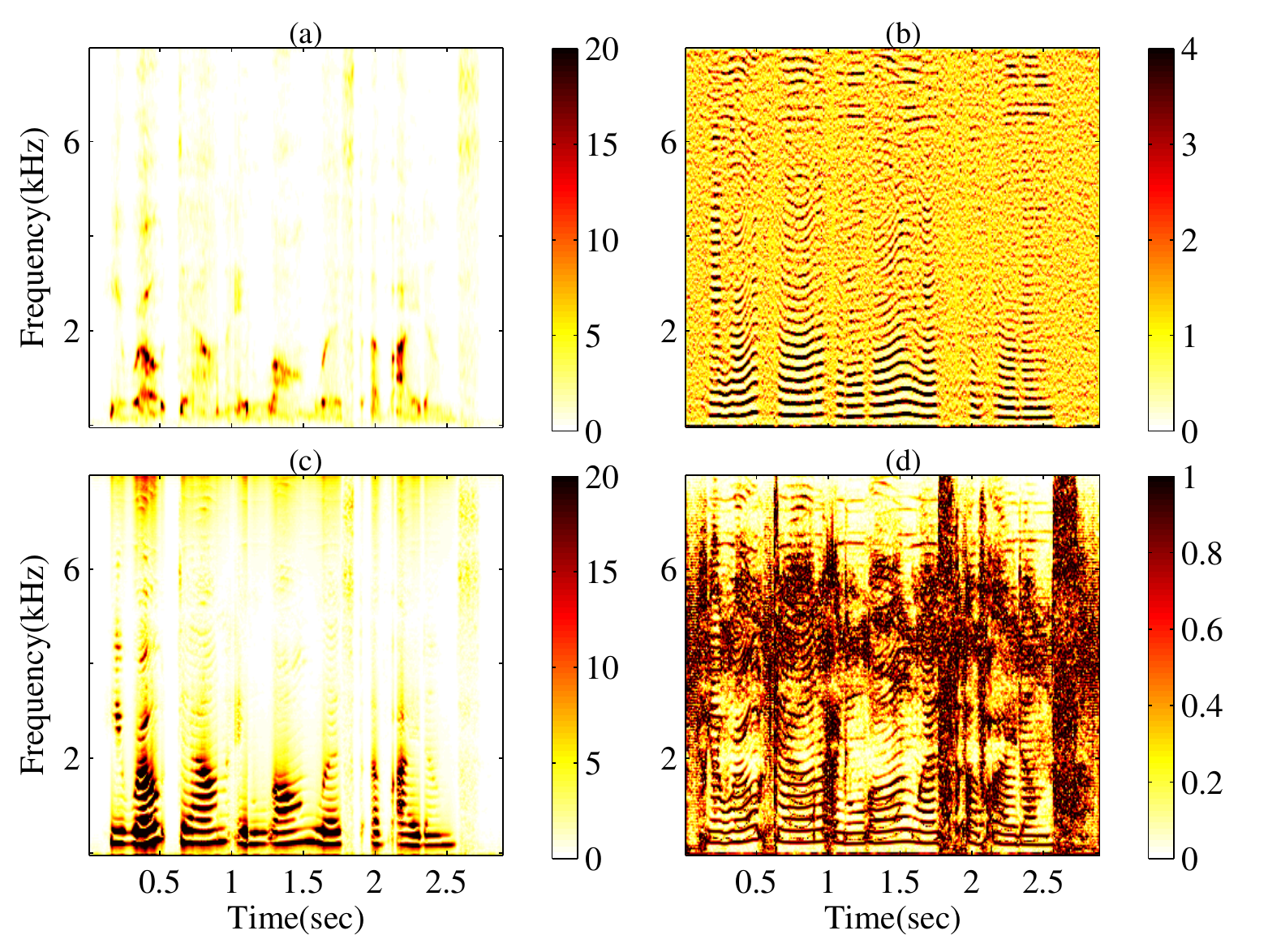}}
\vspace{-4mm}
\caption{The spectrogram of spectral envelop subspace (a)(c), and spectral details subspace (b)(d) obtained by CMI and Hilbert transform, respectively.}
\vspace{-3mm}
\label{sped}
\end{figure}

The obtained subspace matrices by two approaches are shown in Fig.~\ref{sped}. Obviously, in the spectral details subspace, the Hilbert transform produces a considerably irregular speech distribution when the speech deviates from an ideal sinusoidal model. Comparatively, the proposed CMI obtains a periodical alignment of speech components.  

\begin{figure}[!hbt]
\vspace{-2mm}
\centerline{\includegraphics[scale=0.45]{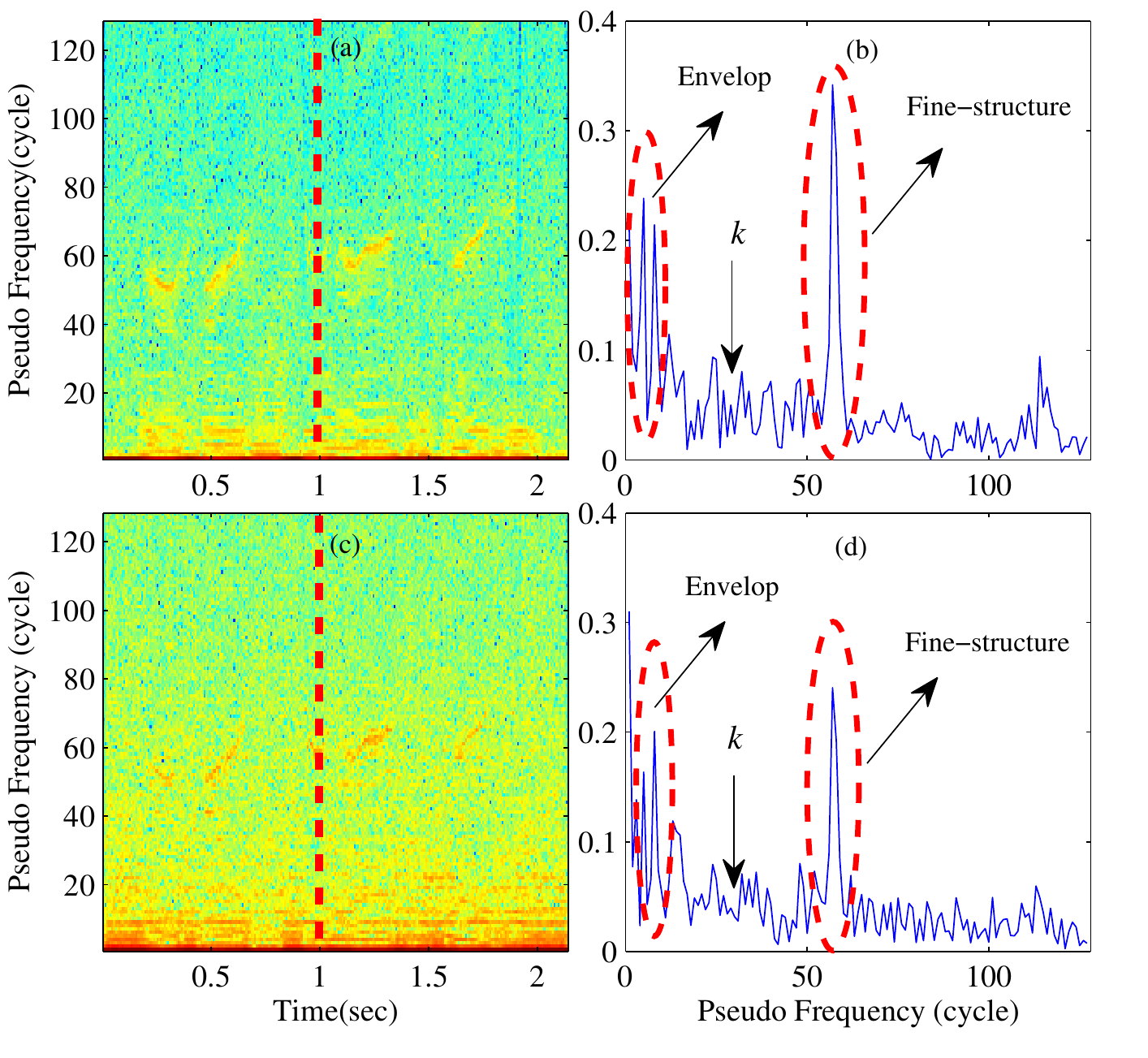}}
\vspace{-4mm}
\caption{The cepstrograms of clean speech (a) and noisy speech (c). Two column slices as indicated by the red lines represent the pseudo frequency components of clean speech (b) and noisy speech (d) at 1 Sec, respectively.}
\label{cped}
\end{figure}

In CMI, the masking matrix $I_{k}$ decides the energy distribution of $Y_{\mathit{e}}$ and $Y_{\mathit{d}}$. To effectively extract the speech components from the two subspaces, $k$ should be optimized to achieve noise trade-off between the two subspaces. As shown in Fig.~\ref{cped}, two peak magnitudes in the cepstrum domain referred as the spectral envelop and spectral details (i.e., fine structure) are located at low pseudo frequency and high pseudo frequency regions (as marked in Fig.~\ref{cped}), respectively. Different types of noises can be translated into varieties of distribution in cepstral domain: babble noises with the speech-like structure can produces 'fake' peaks, and steady Gaussian noises with flat spectral envelop present strong low pseudo frequency peak.  

Along the pseudo frequency axis, $k$ defines SNR in each subspace. According to the Parseval's theorem, we calculate SNR in cepstral domain instead of T-F domain. Therefore, we define SNR$_{e}(k)$=E$(\mathcal{X}_{e}^{2})$/E$(\mathcal{N}_{e}^{2})$ = $\sum_{f^{'}=1}^{k}\mathcal{X}^{2}(f^{'})$/$\sum_{f^{'}=1}^{k}\mathcal{N}^{2}(f^{'})$. \(\mathcal{X}\) and \(\mathcal{N}\) are the speech and noise components in the cepstral domain. Accordingly, SNR$_{d}(k)$ = $\sum_{f^{'}=k+1}^{[N/2]}\mathcal{X}^{2}(f^{'})$/$\sum_{f^{'}=k+1}^{[N/2]}\mathcal{N}^{2}(f^{'})$, where $[\cdot]$ is a round operator. Supposing speech components are ideally concentrated in two narrow bands (as shown in Fig.~\ref{cped}), which are centered at $k_{e}$ and $k_{d}$ with bandwidth $b_{e}$ and $b_{d}$, respectively. To ensure that the major energy of envelop and fine-structure can be separated, $k$ should be in the range [$k_{e}$ $k_{d}$]. Moreover, SNR$_{e}(k)$ can be approximated as $\sum_{k_{e}-b_{e}/2}^{k_{e}+b_{e}/2}\mathcal{X}^{2}(f^{'})$/$\sum_{1}^{k}\mathcal{N}^{2}(f^{'})$ = $E_{e}/\sum_{1}^{k}\mathcal{N}^{2}(f^{'})$. When $k$ increases, $\sum_{1}^{k}\mathcal{N}^{2}(f^{'})$ increases, and accordingly, SNR$_{e}$ will decrease. Thus to obtain higher SNR$_{e}$, $k$ therefore should be as smaller as possible, however, higher than $k_{e}+b_{e}/2$ to maintain the spectral envelop subspace energy. Typically, the vocal range (i.e., fundamental frequency) for human speech is about 85 Hz to 300 Hz \cite{elliott2009modulation}. Accordingly, the lower boundary of $k$ should include 300 Hz fundamental frequency. Hence, we have relationship satisfying 
\begin{equation}
85 \leq f = \frac{F_{s}}{2k} \leq 300
\label{frequency} 
\end{equation}
The sampling frequency $F_{s}$ is 16 kHz, and alternatively, $ 26.6 \leq k$ cycle in half second. In practice, when $k$ is selected as 25-35 cycle, the results are comparable. In our study, $k$ is set as 30.

\subsection{Low Rank and Sparse Characteristics of Two Subspaces}
In the last section, we have demonstrated that with different $k$, CMI can lead to different spectral subspaces. When $k$ is varying between the envelop peak and details peak along the pseudo frequency axis in cepstrum domain, SNRs of $Y_{e}$ and $Y_{d}$ are changed. With a given $k$ within the range in (\ref{frequency}), CMI procedure is equal to the transform $G = W^{-1}\hat{I}_{k}W$, where 
\[
\hat{I}_{k} =
\begin{bmatrix}
I_{k+1}  & 0  &\dots  &0 \\
0        &0   &\dots  &0\\
\dots      & \ddots  &\ddots &\dots\\
0       &0   &\dots &I_{k}\\
\end{bmatrix} \] 
$G$ is a circulant matrix, $G_{i,i}\stackrel{i\neq j}{>} G_{i,j}$, and $\sum_{j}G_{i,j} = 1$. In what follows, the row and column indices $1\leq \alpha \leq n$ and $1\leq \beta \leq m$. Then, the $(\alpha, \beta)$-th entry of $X_{e}$ is given by 
\begin{equation}
(X_{e})_{\alpha, \beta} = [exp(Glog(X))]_{\alpha, \beta} = \prod_{\gamma=1}^{n}X_{\gamma,\beta}^{G_{\alpha,\gamma}}
\label{xes}
\end{equation}
Similarly, by $Y = X+ N^{'}$, 
\begin{equation}
\begin{split}
(N_{e})_{\alpha, \beta} & = \prod_{\gamma=1}^{n}(X_{\gamma,\beta}+N_{\gamma,\beta}^{'})^{G_{\alpha,\gamma}}- \prod_{\gamma=1}^{n}X_{\gamma,\beta}^{G_{\alpha,\gamma}}\\
& = \prod_{\gamma=1}^{n}X_{\gamma,\beta}^{G_{\alpha,\gamma}}\left(\prod_{\gamma=1}^{n}(1+\frac{N_{\gamma,\beta}^{'}}{X_{\gamma,\beta}})^{G_{\alpha,\gamma}}-1\right)
\end{split}
\label{nes}
\end{equation}

\begin{figure}[!hbt]
\center
\vspace{-5mm}
\subfigure{\includegraphics[scale=0.50]{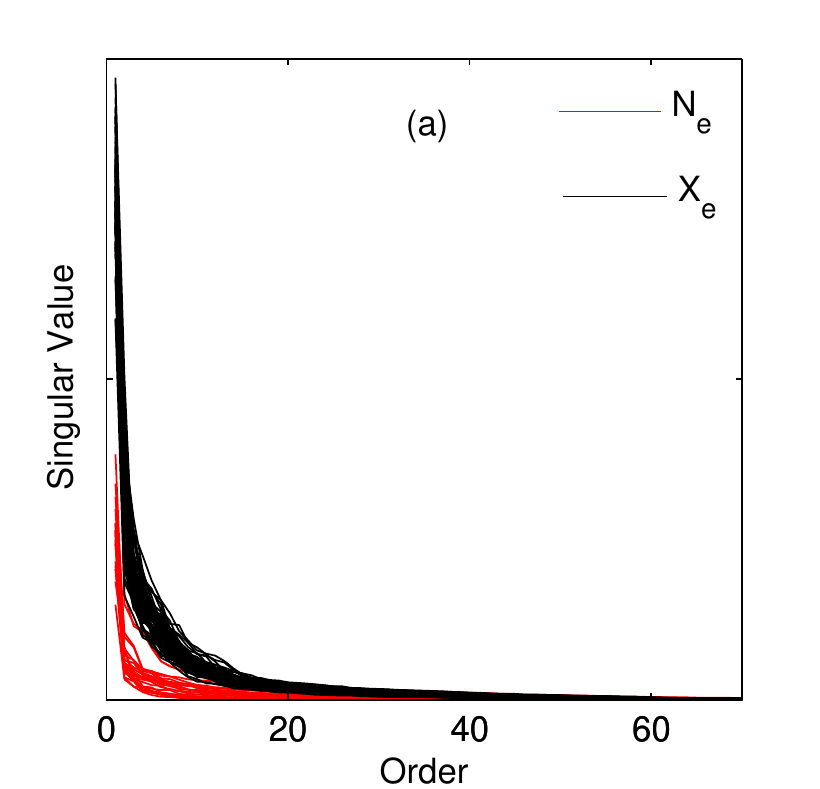}}
\hspace{-2.4em}
\subfigure{\includegraphics[scale=0.50]{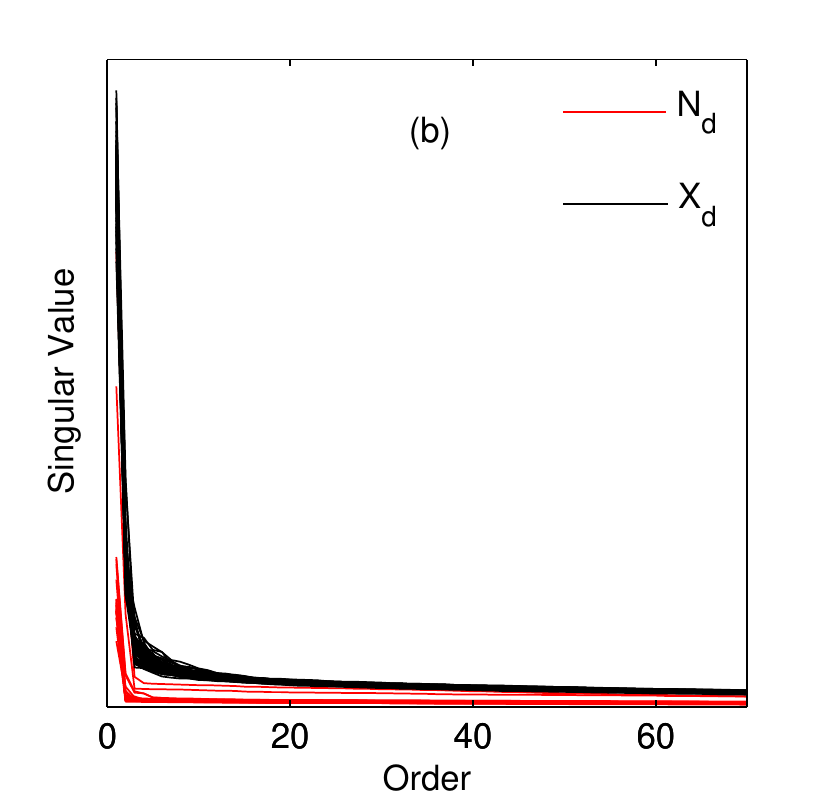}}
\vspace{-5mm}
\caption{The singular values along the diagonal order in envelop subspace (i.e., $N_{e}$, and $X_{e}$) (a) and detail subspace (i.e., $N_{d}$, and $X_{d}$)(b). Here the $k$ is set as 20, and SNR is 0 dB. 30 noise samples and 100 speech samples are used for the experiment, and each line represents the diagonal singular values of one sample in T-F domain.}
\vspace{-3mm}
\label{slowrank}
\end{figure}

$X_{d}$ and $N_{d}$ are in the same form as (\ref{xes}) and (\ref{nes}), in which $G^{'} = W^{-1}(I-\hat{I}_{k})W$ is used to replace $G$. Technically, to prove that $X_{e}$ is relatively higher rank than $N_{e}$ is almost impossible. However, due to the fact that small values in these matrics have no significant impact on speech components recovery, we can focus on these principle components that most influential to noise cancellation. Thereby, singular value decomposition (SVD) is utilized to demonstrate the approximate rank properties. The numerical results have been shown in Fig.\ref{slowrank} 

The experimental results clearly demonstrate that the singular values of noise matrices (i.e., $N_{e}$ and $N_{d}$) decrease faster than speech matrices (i.e., $X_{e}$ and $X_{d}$). By cutting off small singular values (i.e., thresholding by $\epsilon$), both $N_{e}$ and $N_{d}$ show lower rank than $X_{e}$ and $X_{d}$, respectively. This conclusion further can be used to justify the implementation of low-rank decomposition.

\begin{figure}[!hbt]
\vspace{-2mm}
\centerline{\includegraphics[scale=0.55]{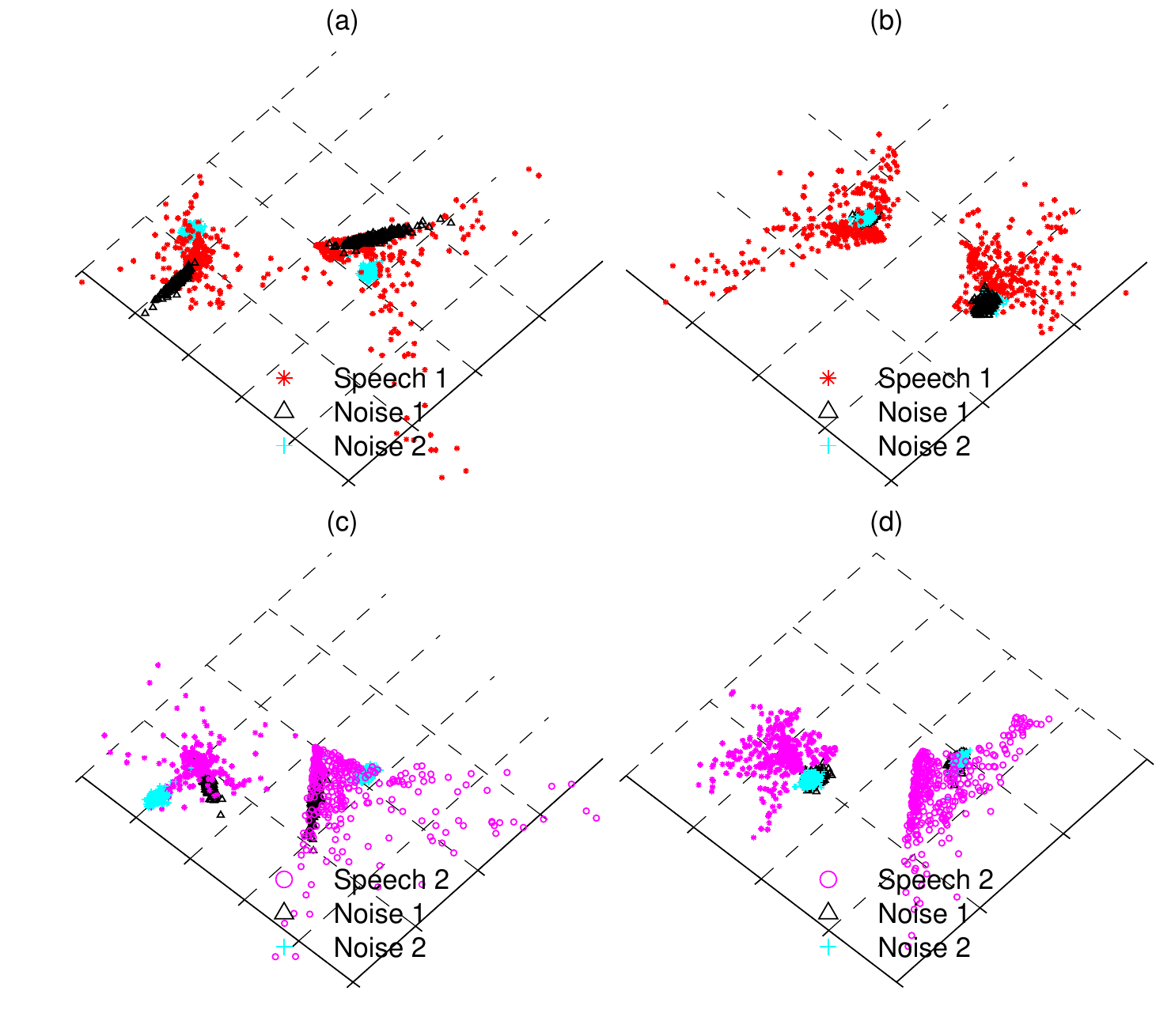}}
\vspace{-4mm}
\caption{The principal components projection of speech and noise samples in the spectral envelop subspace (a) and (c), and in the spectral details subspace (b) and (d), respectively.}
\label{speol}
\end{figure}

In sparsity perspective, speech components generally are more spectrally diverse than noise components \cite{chen2013speech}. This conclusion is also applicable in the proposed two modulation subspaces. The spectral envelop subspace reflects the low pseudo frequency components, and apparently it is more 'smooth' than the spectral details subspace. Such 'smooth' can be regarded as less spectral basis diversity. Specifically, $X_{\mathit{e}}$ and $N_{\mathit{e}}$ are both included in this 'smooth' subspace, which means their spectral convex hulls would be overlapped with high ratio. This intuitive assumption has been evidently shown in Fig.~\ref{speol}, where the speech and noise components are projected to the principal axes (i.e., eigenvectors). The principal directions are extracted from the clean speech spectra, and the first 3 largest and the succeeded 3 secondary largest eigenvectors are used as the 3 dimensional support basis displayed in the same space. As shown in Fig.~\ref{speol}(a)(c), in spectral envelop subspaces, speech bases are severely overlapped, compared with that in the spectral details subspace noise bases are concentrated to be easily separated from the speech bases shown in Fig.~\ref{speol}(b)(d). Therefore, it is easier to separate speech components from noise components in the spectral details domain than in the spectral envelop domain. 

Based on the energy distribution in the cepstrum domain, as the discussion of SNR$_{e}$ and SNR$_{d}$ in II.A, the spectral envelop subspace demonstrates higher SNR compared with the spectral details subspace. The highly overlapped spectral bases between speech and noise components in the spectral envelop subspace require supervised decomposition approach. Contrarily, in the spectral details subspace, speech components can be considered as the summation of several harmonics within narrow frequency band, while the noise components in this subspace generally are random statics. As a result, a general RPCA based decomposition in the spectral details subspace can be used to separate speech and noise components.

\subsection{Noise in the Spectral Envelop Subspace}
\begin{figure}
\center
\vspace{1mm}
\subfigure{\includegraphics[scale=0.25]{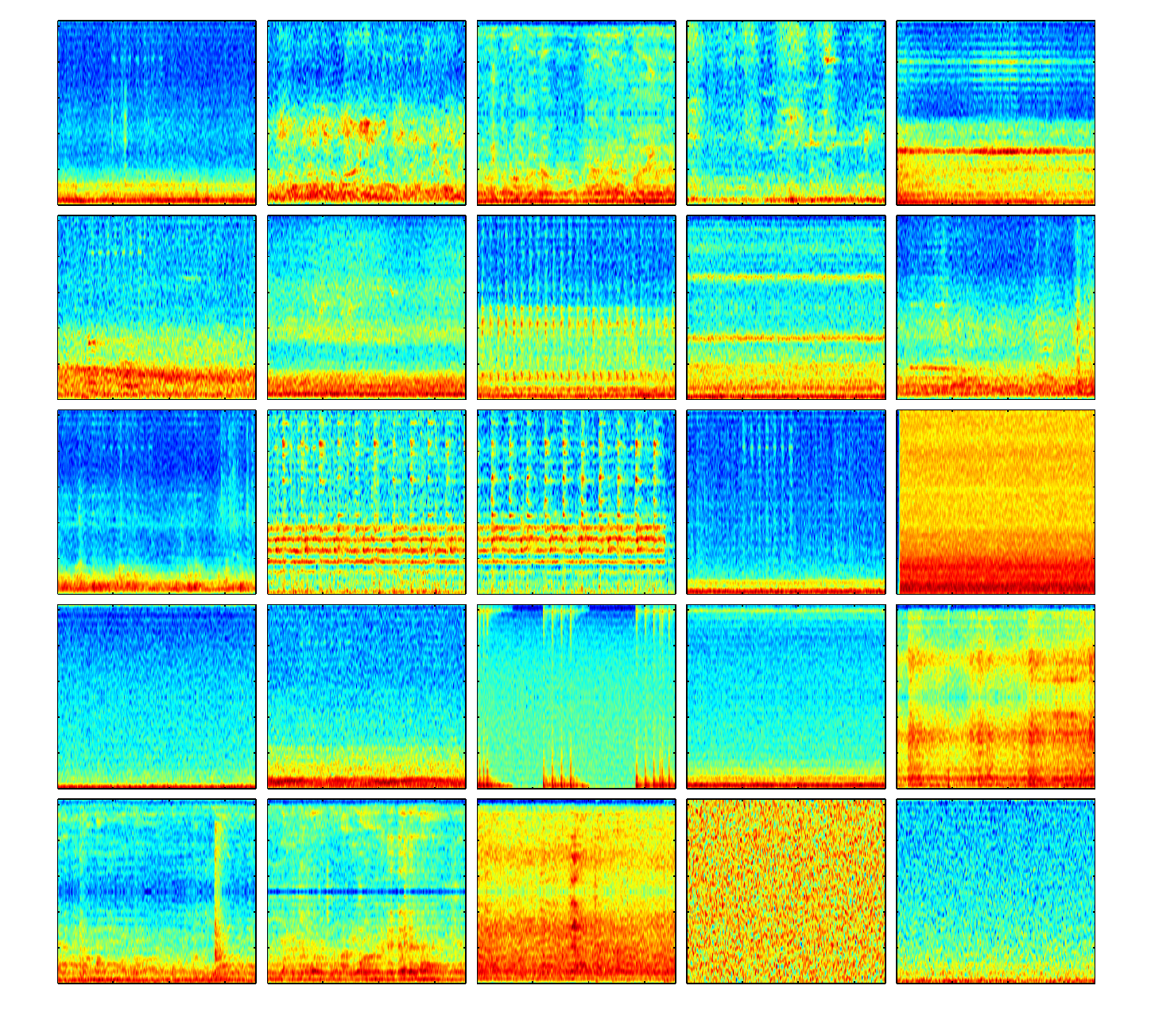}}
\subfigure{\includegraphics[scale=0.31]{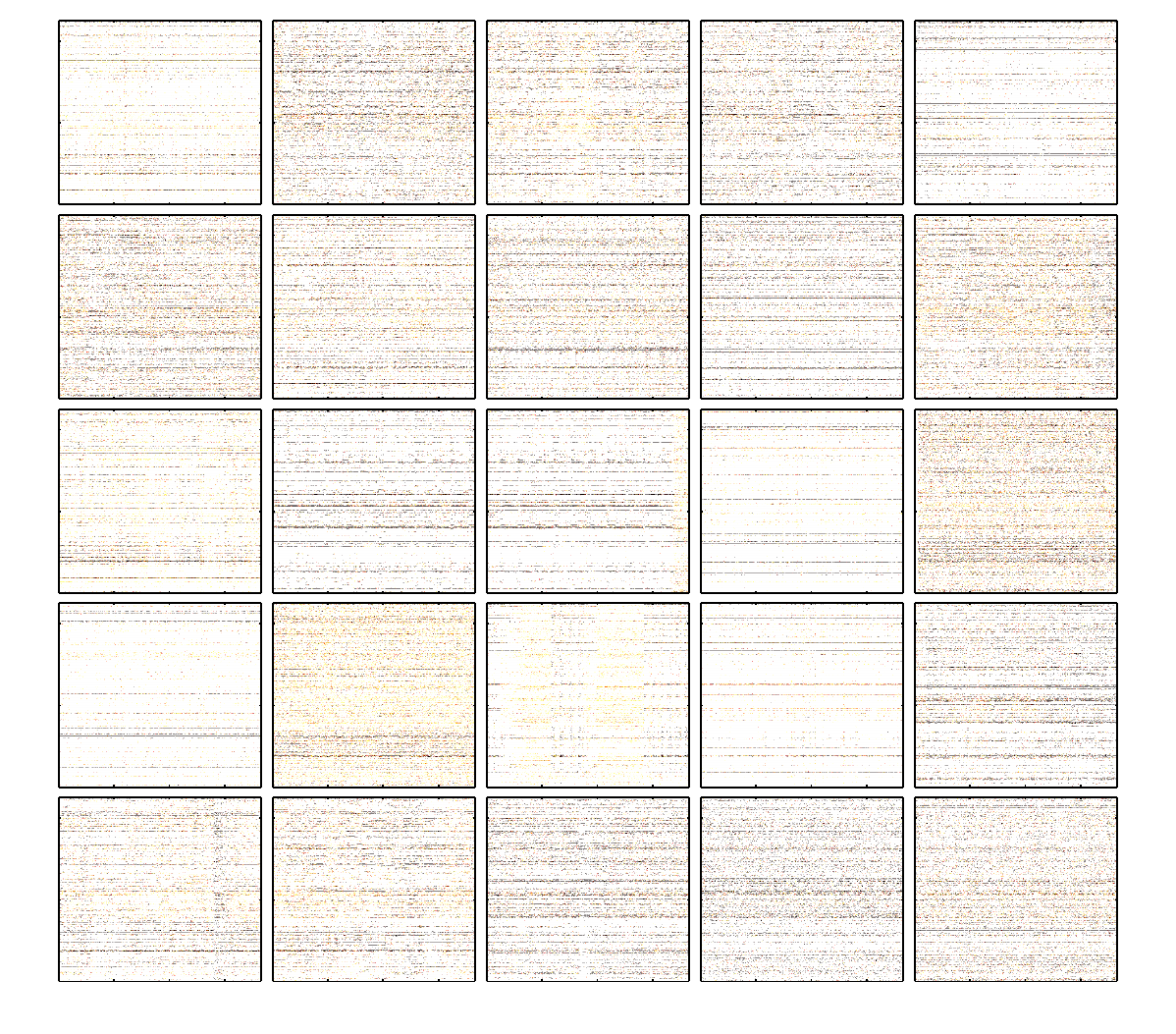}}
\vspace{-3mm}
\caption{The 25 noise samples in the spectral envelop subspace $N_{e}$ (top panel) and in the activation subspace $L_{a}$ of speech dictionary (bottom panel), respectively.}
\vspace{-5mm}
\label{snoise}
\end{figure}

As discussed in II.B, noise components share more bases with speech components in the spectral envelop subspace than in the spectral details subspace. Therefore, we further investigated how the low-rank and sparse characteristics of different types of noise affect the separation of speech and noise components in the spectral envelop subspace. In this study, 25 noise samples (as listed in Table.~\ref{snoise1}) obtained from several databases, including NOISEX-92, IEEE database, and NOIZEUS, are used. The spectrograms of these noise samples are decomposed into the spectral envelop subspace as shown in Fig.~\ref{snoise} (left), and noted as $N_{e}$. For each noise sample, $N_{e}$ has an approximation form consisted of linear combinations of speech bases in the spectral envelop subspace. This projection is given as a non-negative least square optimization  
\begin{equation}
\begin{split}
\min_{L_{a}} \qquad  &\|N_{e}-D_{e}L_{a}\|_{F}^{2} \\
s.t. \qquad  &L_{a}\geq 0
\end{split}
\end{equation}
where the activation matrix $L_{a}$ (as shown in Fig.~\ref{snoise}(right)) is a linear transformation of noise matrix $N_{e}$ into the speech dictionary space $D_{e}$. Both $N_{e}$ and $L_{a}$ matrices show low-rank and sparse characteristics. It means when the LSD is conducted, there is a trade-off on distributing the two parts of noise components into speech subspace.   

To quantitatively describe the impact of different noises on speech enhancement, two indices, referred as coherent ratio and sparsity-to-low-rank ratio (SLR), are proposed to explain the general selection criterion of decomposition algorithm and parameters. The coherent ratio $C$ is applied to measure the coherence between the noise components and speech components, and is given as
\begin{equation}
C(N_{e}, D_{e}) = \sum_{jk} \frac{|\left\langle N_{e,j},D_{e,k}\right\rangle|}{\|N_{e,j}\|_{2}\|D_{e,k}\|_{2}} 
\end{equation}
where $N_{e,j}$ denotes the $j$th column in $N_{e}$ and $D_{e,k}$ is the $k$th column in $D_{e}$. Two matrices are considered less coherent if $C$ is small. The normalized $C$ values of 25 noise samples are summarized in Table.~\ref{snoise1}, in which babble noise is strongly coherent to speech dictionary (\(C\) values at 0.85), and white Gaussian noise is much less coherent to speech dictionary ($C$ values at 0.35). In order to minimize the errors caused by incorrect distributions of noise components, the SLR R is proposed for a better understanding of the noise's energy distribution on low-rank components and sparse components. Accordingly, R is described as
\begin{equation}
R = \frac{\|\cdot\|_{1}}{\|\cdot\|_{*}}
\end{equation}
where $\ell_{1}$ norm is defined as \(\|\cdot\|_{1} = \sum_{ij}|\cdot|_{ij}\). The nuclear norm is $\|\cdot\|_{*} = \sum_{i}\sigma_{i}(\cdot)$, in which $\sigma_{i}(\cdot)$ is the singular value. We calculate the R values of noise samples in both the spectral envelop subspace $N_{e}$ and the activation matrix $L_{e}$, corresponding to $R_{noise} = \|N_{e}\|_{1}/\|N_{e}\|_{*}$ and $R_{Activation} = \|L_{a}\|_{1}/\|L_{a}\|_{*}$, respectively. 

The normalized values of $R$ (as summarized in Table.~\ref{snoise1}) ranging from 0 to 1 indicate which part of the matrix takes superiority in RPCA based decomposition. R$_{Noise}$ represents the R values in envelop subspace, and R$_{Activation}$ represents the R value in speech dictionary space. Different types of noise have various R values, which indicate the degree that they can be decomposed into speech subspace. For instance, Gaussian noise shows the highest sparsity (R$_{Noise}$ = 1) in the spectral envelop subspace, and a considerably low rank (R$_{Activation}$ = 0.06) in the speech dictionary space. In contrast, Volvo noise has the highest sparsity (R$_{Activation}$ = 1) in speech dictionary space, but it is relatively low-rank (R$_{Noise}$ = 0.39) in the spectral envelop subspace. Generally, with the assumption that speech shows sparsity characteristic, one can propose different decomposition tactic strategies to conform the SLR. To develop robust LSD, the consideration of R value of noise can facilitate the separation of speech and noise components in the spectral envelop subspace. 
\begin{center}
\begin{table}[ht]
\caption{The sparsity-low-rank ratio and coherent ratio in (16)(17) are presented. All 25 noise samples are used, and the results are normalized to the range [0, 1].} 
\centering 
\begin{threeparttable}
\begin{tabular*}{0.51\textwidth}{ l c c c} 
\hline 
&\footnotesize C$_{Coherent}$  &\footnotesize R$_{Noise}$ &\footnotesize R$_{Activation}$   \\ \hline
Car 60mph$^{*}$                  &0.57      &0.31       &0.34        \\
Cafeteria babble$^{*}$           &0.82      &0.54       &0.07        \\
Babble$^{\#}$                    &0.85      &0.55       &0.11       \\ 
Construction Crane$^{*}$         &0.81      &0.54       &0.02        \\
Inside Flight$^{\ddagger}$       &0.53      &0.44       &0.66         \\
Street Downtown$^{*}$            &0.79      &0.56       &0.13         \\
Street$^{\ddagger}$              &0.81      &0.56       &0.31         \\ 
Construction Drilling$^{*}$      &0.86      &0.58       &0.21         \\
F16$^{\ddagger}$                 &0.83      &0.92       &0.44        \\
Inside Train 1$^{*}$             &0.84      &0.53       &0.15         \\
Inside Train 2$^{*}$             &0.68      &0.22       &0.39        \\ 
Train1$^{\#}$                    &0.40      &0.75       &0.55         \\ 
Train2$^{\ddagger}$              &0.22      &0.62       &0.52         \\ 
PC Fan$^{*}$                     &0.27      &0.31       &0.84        \\
SSN$^{*}$                        &1         &0.69       &0.22         \\
Volvo$^{\ddagger}$              &0.01      &0.39       &1            \\
Water Cooler$^{*}$               &0.86      &0.56       &0.30        \\
Machinegun$^{\ddagger}$          &0.52      &0.01       &0.01        \\ 
Leopard$^{\ddagger}$             &0.36      &0.31       &0.81        \\ 
Subway$^{\#}$                    &0.84      &0.85       &0.26        \\
Airport$^{\#}$                   &0.77      &0.47       &0.16        \\
Restaurant$^{\#}$                &0.85      &0.54       &0.16        \\
Exhibition$^{\#}$                &0.74      &0.89       &0.21        \\
White Gaussian$^{\$}$            &0.35      &1          &0.06        \\
Pink$^{\$}$                      &0.85      &0.87       &0.23        \\
\hline 
\end{tabular*}
\begin{tablenotes}
\item[$*$] IEEE; \item[$\ddagger$] Noisex 92; \item[$\#$] NOIZUS; \item[$\$$] Simulated
\end{tablenotes}
\end{threeparttable}
\label{snoise1} 
\end{table}
\end{center}

\section{Semi-Supervised Decomposition in modulation subspace}
In this study, two different supervised LSD schemes are proposed in the spectral envelop subspace, and both of them are applied to obtain a robust recovery of speech components. In the spectral details subspace, a standard unsupervised RPCA is implemented.

\subsection{Model Description}
In both spectral envelop and spectral details subspaces, LSD is conducted. A basic RPCA model for both subspaces can be written as
\begin{equation}
\begin{split}
\min_{X_{\dagger}} \qquad  & \| X_{\dagger}\|_{1}+\lambda\|N_{\dagger}\|_{*} \\ 
s.t. \qquad & Y_{\dagger}  = X_{\dagger}+N_{\dagger}
\end{split} 
\end{equation}
where $\dagger \in \{e,d\}$ refers to the spectral envelop subspace and the spectral details subspace, respectively. As discussed in Section II, noise components share more bases with speech components in the spectral envelop subspace than that in the spectral details subspace. In other words, it is harder to distinguish noise from speech in the spectral envelop subspace without prior information. Therefore, the offline-trained speech dictionary $D_{e}$ in the spectral envelop subspace is employed to separate $N_{e}$ from $X_{e}$.
\subsubsection{spectral envelop subspace model}
In a supervised decomposition, $X_{e}$ is commonly regarded as a sparse activation of a global speech dictionary. The underlying reason is that speech components in each local segment can be expressed by a few bases within a global speech dictionary space. A sparse constrain can be enforced upon the activation matrix to extract the speech components. Comparatively, for $N_{e}$, the low-rank constraints implemented in the spectral envelop subspace can effectively pick out noise components which are not within the speech dictionary space, alternatively incoherent to speech components. However, the noise components that are coherent to speech dictionary bases may not be excluded from speech components. As a result, the activation matrix of speech dictionary is mixed with some outlying entries, which represents the noise approximation. As discussed in Section II, when the SLR varies, the noise components show different probabilities to be decomposed into the speech subspace. Therefore, in this study, we proposed two different supervised decompositions, referred as a two-layers LSD (TLSD-MS) and a single layer LSD (SLSD-MS), in the spectral envelop subspace. Both decompositions are based on the consideration of reducing incoherent noise components.    

The proposed TLSD-MS is straightforward on incoherent noise cancellation: after first layer LSD in the spectral envelop subspace, the second layer LSD is implemented in the activation matrix of speech dictionary. The first layer LSD can be written as    
\begin{equation}
\begin{split}
\min_{S_{1}} \qquad  & \| S_{1}\|_{1}+\lambda_{L_{e,2}}\|L_{e,2}\|_{*} \\
s.t. \qquad & Y_{e}  = D_{e}S_{1}+L_{e,2}
\end{split}
\label{problem1}
\end{equation}
where the spectral envelop matrix $Y_{e}$ is decomposed as the summation of a low-rank matrix $L_{e,2}$, reflecting less spectrally diverse noise components, and a product of speech dictionary $D_{e}$ and its sparse activation matrix $S_{1}$  \cite{candes2011robust}. In the speech components $D_{e}S_{1}$, it either exists some noise residuals or is highly distorted by tightening or relaxing the noise constraint coefficient $\lambda_{L_{e,2}}$. A natural thought is to conduct the second layer LSD in the obtained $S_{1}$. Accordingly, it can be translated as the following optimization problem
\begin{equation}
\begin{split}
\min_{S_{e}} \qquad  & \| S_{e}\|_{1}+\lambda_{L_{e,1}}\|L_{e,1}\|_{*} \\
s.t. \qquad & S_{1}  = S_{e}+L_{e,1}
\end{split}
\label{problem2}
\end{equation}
where $L_{e,1}$ is the activation matrix of noise components in speech dictionary space. The rationality of (\ref{problem2}) is that even in the activation matrix, speech components $S_{e}$ are still more diverse than noise components $L_{e,1}$, which has been explained in the Fig.~\ref{speol}(a) and (c). The TLSD-MS is quite efficient when the noise components have small R$_{noise}$ and large R$_{activation}$ (e.g., volvo noise listed in Table~\ref{snoise1}). 

When the background noise shows large R$_{noise}$, the first layer of TLSD-MS may decompose the noise components into speech with a high possibility. Moreover, the second layer of TLSD-MS only works well for those noise with high R$_{activation}$ in the speech dictionary space. Hence, another decomposition scheme SLSD-MS is proposed as  
\begin{equation}
\begin{split}
\min_{S,L} \qquad  & \| S_{e}\|_{1}+\lambda_{L_{e,1}}\|L_{e,1}\|_{p}+\lambda_{L_{e,2}}\|L_{e,2}\|_{*} \\
s.t. \qquad & Y_{e}  = D_{e}(S_{e}+L_{e,1})+L_{e,2}
\end{split}
\label{problem3}
\end{equation}
where $L_{e,1}$ and $L_{e,2}$ have the same definitions as in TLSD-MS, reflecting the coherent and incoherent noise components, respectively. $\|\cdot\|_{p} = \sum f(\sigma(\cdot))$ is a modified nuclear norm, in which $\sigma(\cdot)$ is the singular value, and $f$ is a mapping function, defined as $f(t) = \frac{(1+p)t}{p+t}$ \cite{kang2015robust}. The conventional nuclear norm over-penalizes large singular values, and consequently may only find a biased solution. In the p-type norm, when $p \rightarrow 0$, a tighter rank approximation can be obtained. Specifically, this tight-rank approximation $\|\cdot\|_{p}$ can equally treat each singular value in the optimization.

\subsubsection{spectral details subspace model}
In the spectral details subspace, the typical RPCA decomposition is given by 
\begin{equation}
\begin{split}
\min_{S_{d},L_{d}} \qquad  & \| S_{d}\|_{1}+\lambda\|L_{d}\|_{*} \\
s.t. \qquad & Y_{d}  = S_{d}+L_{d}
\end{split}
\label{problem4}
\end{equation}
where $S_{d}$ is considered as the approximation to $X_{d}$, and $L_{d}$ corresponds to $N_{d}$. This unsupervised LSD can effectively separate speech components from noise components in the spectral details subspace. Because the speech components in the spectral details subspace show periodic structures (i.e., fine-structure as shown in Fig. 2(b)), the conventional RPCA can work well on speech extraction in the spectral details subspace. 

\subsection{Dictionary Learning}
NMF is a reliable method to obtain speech dictionary \cite{gemmeke2011exemplar}, and it can be given as 
\begin{equation}
(D,L) = \argmin_{D,L} D(X\|DL)+\mu r(D,L) 
\label{eqn1}
\end{equation}
where $D(X\|\widehat{X})$ is a cost function, $r(\cdot)$ is an optional regularization term, and $\mu$ is the regularization weight. The minimization of (\ref{eqn1}) is performed under the nonnegativity constraint of $D$ and $L$. The commonly used cost functions include Euclidean distance, Bregman divergence, and the negative likelihood in the probabilistic NMFs. In this study, a Bayesian NMF is used to learn the speech dictionary from the spectral envelop subspace $X$ of clean speech samples. Accordingly, the input matrix is assumed to be stochastic. To perform NMF as $X \approx DL$, the following model is considered:
\begin{equation}
\begin{split}
X_{nm} & = \sum_{i}H_{nim}\\
f_{H_{nim}}(H_{nim}) & = \mathcal{PO}(H_{nim};D_{ni}L_{im}) \\
 & = (D_{ni}L_{im})^{H_{nim}}e^{-D_{ni}L_{im}}/(H_{nim}!)
\end{split}
\end{equation}
where $H_{nim}$ are latent variables, $\mathcal{PO}(h;\lambda)$ denotes the Poisson distribution, and $H!$ is the factorial of $H$. Before conducting the dictionary learning procedure proposed in \cite{mohammadiha2013supervised}, the speech samples are reduced according to the syllable boundaries in the time domain. By this way, it excludes spectral interfere from those utterance interval, which is quite similar with noise structure.

In practice, three approaches (i.e., least angle regression with coherence (LARC), efficient Sparse Coding \cite{lee2006efficient}, and Bayesian NMF) have been applied to learn the clean speech dictionary in the spectral envelop subspace. Among these dictionaries learned by the three approaches, the Bayesian NMF based dictionary showed the best speech enhancement results. Therefore, Bayesian NMF was selected as the dictionary learning method in this study.

\subsection{Algorithms}
To solve the two-layer optimization problem in (\ref{problem1}) and (\ref{problem2}), augmented Lagrangian method (ALM) is employed. For the first layer, the optimization solution is given as  
\begin{equation}
\begin{split}
&\mathcal{L}(S_{1},A,L_{e,2},\rho,\Delta_{1},\Delta_{2}) \\
&= \|A\|_{1}+\lambda_{L_{e,2}}\|L_{e,2}\|_{*}+\frac{\rho}{2}\|A-S_{1}+\frac{\Delta_{1}}{\rho}\|_{F}^{2}\\
&+\frac{\rho}{2}\|Y_{e}-D_{e}S_{1}-L_{e,2}+\frac{\Delta_{2}}{\rho}\|_{F}^{2}\\         
\end{split}
\label{answer1}
\end{equation}
where an auxiliary variable $A$ is introduced , and assumed to be equal to $S_{1}$. After obtaining $S_{e}$, a standard RPCA based decomposition is implemented in the second layer
\begin{equation}
\begin{split}
&\mathcal{L}(S_{e},L_{e,1},\rho,\Delta_{1}) \\
&= \|S_{e}\|_{1}+\lambda_{L_{e,1}}\|L_{e,1}\|_{*}+\frac{\rho}{2}\|S_{1}-S_{e}-L_{e,1}+\frac{\Delta_{1}}{\rho}\|_{F}^{2}       
\end{split}
\label{answer2}
\end{equation}  

The proposed SLSD-MS presented in (\ref{problem3}) can be solved by the alternative direction method of multipliers (ADMM) or ALM. However, both methods require introducing two auxiliary variables to solve (\ref{problem3}) and expensive matrix inversions are required in each iteration. Accordingly, a recently developed method called the linearized alternating direction method with adaptive penalty (LADMAP) is applied in this study. The augmented Lagrangian function is 
\begin{equation}
\begin{split}
&\mathcal{L}(S_{e},B,L_{e,1},L_{e,2},\rho,\Delta_{1},\Delta_{2}) \\
&= \|B\|_{1}+\lambda_{L_{e,1}}\|L_{e,1}\|_{p} \\
&+\lambda_{L_{e,2}}\|L_{e,2}\|_{*}+\frac{\rho}{2}\|B-S_{e}+\frac{\Delta_{1}}{\rho}\|_{F}^{2}\\
&+\frac{\rho}{2}\|Y_{e}-D_{e}(S_{e}+L_{e,1})-L_{e,2}+\frac{\Delta_{2}}{\rho}\|_{F}^{2}\\         
\end{split}
\label{answer3}
\end{equation}
where the introduced extra variable $B$ is assumed to be equal with $S_{e}$. Specifically, to update $L_{e,1}$, a subproblem is proposed as  
\begin{equation}
\begin{split}
\min_{L_{e,1}}  \lambda_{L_{e,1}}\|L_{e,1}\|_{p}+ \frac{\rho}{2}\|Y_{e}-D_{e}(S_{e}+L_{e,1})-L_{e,2}+\frac{\Delta_{2}}{\rho}\|_{F}^{2}\\       
\end{split}
\label{answer3_1}
\end{equation}
here we extend the method proposed in \cite{kang2015robust} to LADMAP approach. Therefore, the conventional singular value thresholding operator can be redefined as 
\begin{equation}
\Theta_{\partial f(\sigma)\tau}^{'}(Y) = U\mathcal{S}_{\partial f(\sigma)\tau}^{'}(\Sigma)V^{*}
\label{answer3_2}
\end{equation}
where \(Y = U\Sigma V^{*}\) is any singular value decomposition and \(\sigma \in \Sigma\). Considering the thresholding value \(\partial f(\sigma)\tau\) includes the singular value \(\sigma\) itself, an iterative approach is applied to yield the converged \(\sigma\). Accordingly, the shrinkage operator \(\mathcal{S}_{\partial f(\sigma)\tau}^{'}(\Sigma)\) has a closed-form solution at the \(j\)th inner iteration
\begin{equation}
\sigma^{j+1} = (\sigma_{\Sigma}-\partial f(\sigma^{j})\tau)_{+}
\end{equation}
 
For (\ref{problem4}), a same solution form as (\ref{answer2}) can be proposed as
\begin{equation}
\begin{split}
&\mathcal{L}(S_{d},L_{d},\rho,\Delta_{1}) \\
&= \|S_{d}\|_{1}+\lambda_{L_{d}}\|L_{d}\|_{*}+\frac{\rho}{2}\|Y_{d}-S_{d}-L_{d}+\frac{\Delta_{1}}{\rho}\|_{F}^{2}    
\end{split}
\label{answer4}
\end{equation} 

As a relaxation of \(\ell_{0}\), the \(\ell_{1}\) norm is the summation of the absolute values of all the entries. Therefore, it may lead to suppression of speech components, since speech components generally present high intensity in activation matrix compared with noise. To tight the sparsity constraint, a special strategy w.r.t energy concentration is proposed for (\ref{answer1})-(\ref{answer4}), and can be given as 
\begin{equation}
S_{\dagger} = max(S_{\dagger},\theta)      
\end{equation}
where \(\dagger \in \{1,e,d\}\). The \(\theta\) is an energy threshold value set for concentrating the decomposition. The introduced energy threshold is a penalty to the sparsity relaxation, and also helps to distinguish the speech components from noise components in the activation matrix.

Generally, (\ref{answer1}), (\ref{answer2}), and (\ref{answer4}) can be solved by ALM, and (\ref{answer3}) is solved by LADMAP. With some algebra, the corresponding updating schemes of TLSD-MS and SLSD-MS are outlined in Algorithm 1 and Algorithm 2, respectively.
\begin{algorithm}
\caption{{\bf Proposed model to solve problem (\ref{answer1})} \label{ALM}}
{\bfseries Input:}noisy speech envelop matrix Y$_{e}$ $\in \mathbb{R}^{M\times N}$ and offline-trained dictionary $D_{e}$, parameters $\lambda_{L_{e,2}}>0$, $\rho^{0}>0$, $\theta$, tolerance $\epsilon$ and $\mu>1$. \\
{\bfseries Initialize:}Set maxIter, and Terminate $\leftarrow$ False. Initialize $S_{1}^{0}$, $A^{0}$, $L_{e,2}^{0}$, $\Delta_{1}^{0}$ and $\Delta_{2}^{0}$ to zero. 
\begin{algorithmic}[1]
\WHILE{Terminate=False}
   \STATE Update $S_{1}^{k+1}$: 
        \begin{align*}
        S_{1}^{k+1}=(D_{e}^{T}D_{e}&+I)^{-1}(D_{e}^{T}Y_{e}-D_{e}^{T}L_{e,2}^{k}\\
        &+A^{k}+\frac{D_{e}^{T}\Delta_{2}^{k}}{\rho^{k}}+\frac{\Delta_{1}^{k}}{\rho^{k}})\\
        S_{1}^{k+1} & = max(S_{1}^{k+1}, \theta)
        \end{align*}
   \STATE Update $A^{k+1}$: 
    \begin{align*}
        A^{k+1} = \max(SR_{1/\rho^{k}}(S_{1}^{k+1}+\frac{\Delta_{1}^{k}}{\rho^{k}}),0)
    \end{align*}
   \STATE Update $L_{e,2}^{k+1}$:
      \begin{align*}
        L_{e,2}^{k+1} = \Theta_{\lambda_{L_{e,2}}/\rho^{k}}(Y_{e}-D_{e}S_{1}^{k+1}+\frac{\Delta_{2}^{k}}{\rho^{k}})
      \end{align*}
\STATE Update the Lagrangian multipliers:
\begin{align*}
\Delta_{1}^{k+1} &= \Delta_{1}^{k}+\rho^{k}(A^{k+1}-S_{1}^{k+1})\\
\Delta_{2}^{k+1} &= \Delta_{2}^{k}+\rho^{k}(Y_{e}-D_{e}S_{1}^{k+1}-L_{e,2}^{k+1})\\
\rho^{k+1} &= \mu\rho^{k}.
\end{align*}
\IF{$\|A^{k+1}-S_{1}^{k+1}\|_{\infty}\leq \epsilon$ and $\|A^{k+1}-A^{k}\|_{\infty}\leq \epsilon$ and $\|L_{e,2}^{k+1}-L_{e,2}^{k}\|_{\infty}\leq \epsilon$ or (k $\geq$ maxIter)}
   \STATE Terminate $\leftarrow$ True\;
\ENDIF
\ENDWHILE
\end{algorithmic}
{\bfseries Output:}Optimal active coefficient matrix $S_{*}=S_{1}^{k}$ \\
\end{algorithm}

\begin{algorithm}
\caption{{\bf Proposed model to solve problem (\ref{answer3})} \label{LADMAP}}
{\bfseries Input:} noisy speech envelop matrix Y$_{e}$ $\in \mathbb{R}^{M\times N}$ and offline-trained dictionary $D_{e}$, parameters $\lambda_{L_{e,1}}>0$ and $\lambda_{L_{e,2}}>0$, $\rho^{0}>0$, $\theta$, and $\mu>1$. \\
{\bfseries Initialize:} Set maxIter, tolerance $\epsilon$ and Terminate $\leftarrow$ False. Initialize $S_{e}^{0}$, $B^{0}$, $L_{e,1}^{0}$, $L_{e,2}^{0}$, $\Delta_{1}^{0}$ and $\Delta_{2}^{0}$ to zero. 
\begin{algorithmic}[1]
\WHILE{Terminate=False}
   \STATE Update $S_{e}^{k+1}$: 
        \begin{align*}
        S_{e}^{k+1}=&(D_{e}^{T}D_{e}+I)^{-1}(D_{e}^{T}Y_{e}-D_{e}^{T}D_{e}L_{e,1}^{k}-\\
        &D_{e}^{T}L_{e,2}^{k}+B^{k}+\frac{D_{e}^{T}\Delta_{2}^{k}}{\rho^{k}}+\frac{\Delta_{1}^{k}}{\rho^{k}}) \\
        S_{e}^{k+1} & = max(S_{e}^{k+1}, \theta)
        \end{align*}
   \STATE Update $B^{k+1}$: 
    \begin{align*}
        B^{k+1} = \max(SR_{1/\rho^{k}}(S_{e}^{k+1}+\frac{\Delta_{1,k}}{\rho^{k}}),0)
    \end{align*}
   \STATE Update $L_{e,1}^{k+1}$:
      \begin{align*}
         L_{e,1}^{k+1} &=\Theta_{\partial f(\sigma)\lambda_{L_{e,1}}(\eta\rho^{k})^{-1}}^{'}(L_{e,1}^{k} + D_{e}^{T}(Y_{e}-D_{e}(S_{e}^{k+1} \\
         &+L_{e,1}^{k})-L_{e,2}^{k}+\frac{\Delta_{2}^{k}}{\rho^{k}})/\eta)
       \end{align*}
    \STATE Update $L_{e,2}^{k+1}$:
      \begin{align*}
        L_{e,2}^{k+1} = \Theta_{\lambda_{L_{e,2}}/\rho^{k}}(Y_{e}-D_{e}(S_{e}^{k+1}+L_{e,1}^{k+1})+\frac{\Delta_{2}^{k}}{\rho^{k}})
      \end{align*} 
\STATE Update the Lagrangian multipliers:
\begin{align*}
\Delta_{1}^{k+1} &= \Delta_{1}^{k}+\rho^{k}(Y_{e}-D_{e}(S_{e}^{k+1}+L_{e,1}^{k+1})-L_{e,2}^{k+1})\\
\Delta_{2}^{k+1} &= \Delta_{2}^{k}+\rho^{k}(B^{k+1}-S_{e}^{k+1})\\
\rho^{k+1} &= \mu\rho^{k}.
\end{align*}
\IF{$\|B^{k}-S_{e}^{k}\|_{\infty}\leq \epsilon$ and $\|B^{k+1}-B^{k}\|_{\infty}\leq \epsilon$ and $\|L_{e,1}^{k+1}-L_{e,1}^{k}\|_{\infty}\leq \epsilon$ or (k $\geq$ maxIter)}
   \STATE Terminate $\leftarrow$ True\;
\ENDIF
\ENDWHILE
\end{algorithmic}
{\bfseries Output:}Optimal active coefficient matrix $S_{*}=S_{e}^{k}$ \\
\end{algorithm}

$\mathcal{SR}_{\tau}$ denote the shrinkage operator $ \mathcal{SR}_{\tau}(x) = sgn(x)max(|x|-\tau, 0)$. and $\Theta_{\tau}(X)$ denote the singular value thresholding operator given by $\Theta_{\tau}(X) = U\mathcal{SR}_{\tau}(\Sigma)V^{*}$, where $X = U\Sigma V^{*}$ \cite{zhuang2012non}. The procedure of algorithm 1 can be implemented for both (\ref{answer2}) and (\ref{answer4}), in which \(S_{e}\) and \(S_{d}\) are updated as \(S_{1}\), and \(L_{e,1}\) and \(L_{d}\) are updated as \(L_{e,2}\). In algorithm 2,  $\Theta^{'}$ is the modified singular value thresholding operator mentioned in (\ref{answer3_2}), in which the shrinkage operator \(\mathcal{S}_{\partial f(\sigma)\tau}^{'}(\Sigma)\) is implemented in an inner iterative until divergence. $\eta=\|Y_{1}\|_{2}^{2}$. The iteration value 'maxIter' is set as 500 in our numerical experiment. The parameter $\mu$ is set as 1.2 to update thresholding value $\rho^{k}$. In the singular value threshold $\frac{\lambda_{L_{e,2}}}{\rho^{k}}$, $\lambda_{L_{e,2}}$ is a SNR related constant. 

\section{Experiments and Results}
In this section, the proposed speech enhancement algorithms, TLSD-MS and SLSD-MS, are evaluated and compared with other state-of-the-art speech enhancement algorithms. In Section IV-A, a direct comparison of the proposed algorithms in modulation subspaces and complete spectrum. In Section IV-B, noise with different coherent ratios are utilized to obtain a better understanding of the merits of TLDS-MS and SLDS-MS algorithms. Evaluations via benchmark metrics and intelligibility indexes are presented in Section IV-C and IV-D, respectively.

In our simulation, all speech and noise signals are down-sampled to 16 kHz and the DFT was implemented using a frame length of 512 samples and 0.5-overlapped Hann windows. We select 600 samples from IEEE database \cite{loizou2013speech} for the noise reduction evaluation. The signal synthesis is performed using the overlap-and-add procedure, and 23 noise samples selected from environmental and industrial noise database \cite{hu2008evaluation} plus two simulated noise (white Gaussian and pink noise) are used at various input SNRs (-10 dB to 10 dB). The dictionaries are all consisted of 750 basis vectors learned from selected 150 speech samples. 

\subsection{Comparison of Proposed algorithms in Modulation Subspaces and Complete Spectrum} 
In order to evaluate the benefits of separated subspace, the proposed TLSD-MS and SLSD-MS are both applied to modulation subspaces and complete spectrum. For the complete spectrum speech recovery, the dictionary is also trained using Bayesian NMF in the T-F domain instead of envelop subspace. The selected 50 speech samples are not overlapped with the test utterances.  White Gaussian, Pink and Volvo are used as the additive background noises at various SNRs (-10, -5, 0, 5, and 10 dB). Four objective metrics, perceptual evaluation of speech quality (PESQ), segmental SNRs (SegSNR) \cite{hu2008evaluation}, signal to distortion (SDR), and hearing-aid speech quality index (HASQI) are used to quantitatively evaluate the performance of the two LSD algorithms in different subspaces.
\begin{figure}[!hbt]
\vspace{-3mm}
\centerline{\includegraphics[scale=0.50]{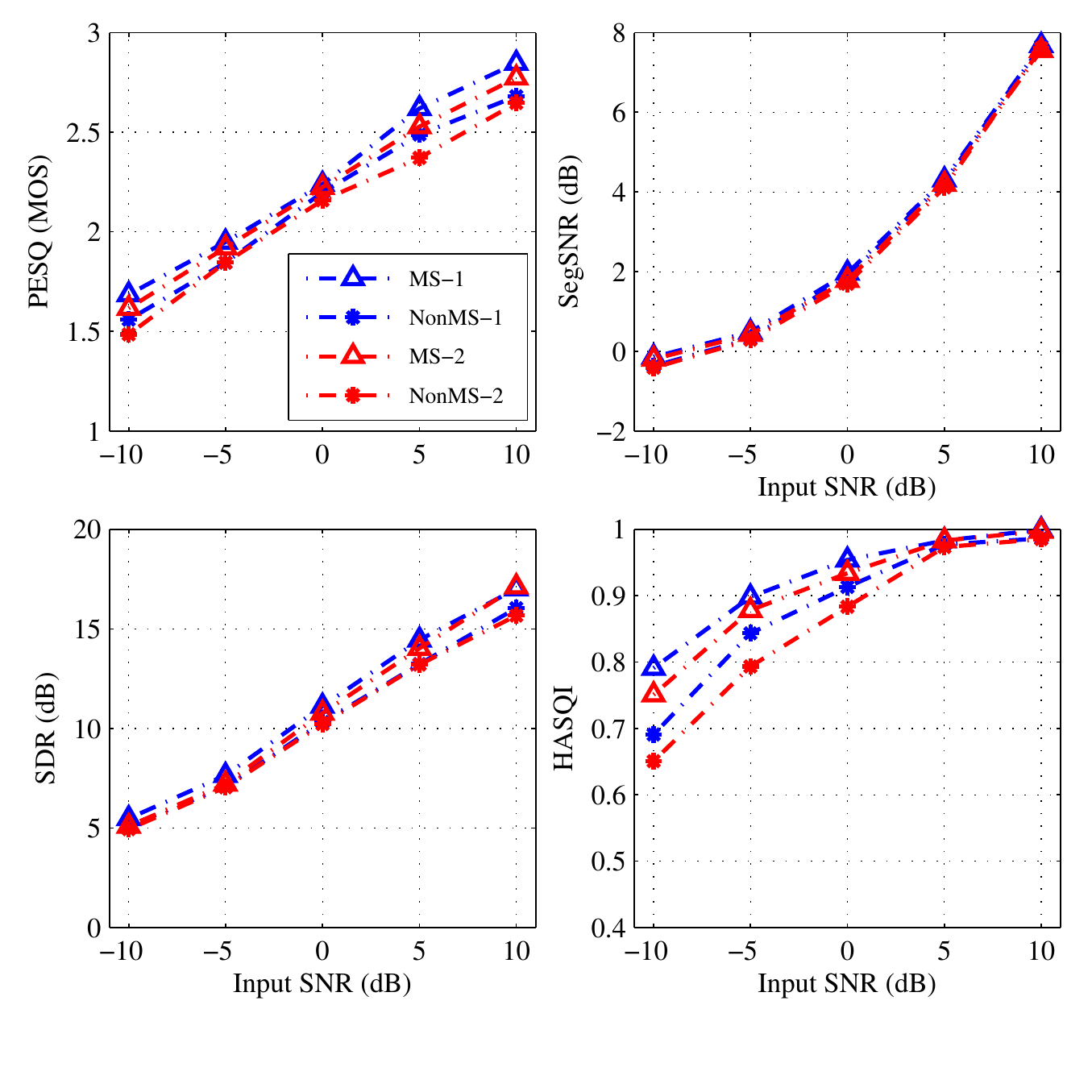}}
\vspace{-7mm}
\caption{Performance evaluations of two proposed algorithms at various SNRs by using different subspaces. In the legend, 'MS' refers to the modulation subspaces based decomposition, and 'NonMS' is the complete spectrum domain based decomposition. The suffix 1 and 2 refer to the algorithm SLSD-MS and TLSD-MS, respectively.}
\vspace{-4mm}
\label{TwoDic}
\end{figure}

As shown in Fig.~\ref{TwoDic}, SLSD-MS generally achieve the similar performances compared with TLSD-MS with respective to PESQ, SegSNRs and SDR metrics. Based on the fine-tune parameters, the two different LSD algorithms show almost the same capability on improving the speech qualities. In terms of speech perception, higher value of HASQI represents better to be recognized with lower error rate. The results clearly demonstrate that the modulation subspace based decomposition (indexed by blue and red $\Delta$) by two algorithms both show higher HASQI values averagely than complete spectrum based decomposition. It indicates that the implementation of decoupling the envelop subspace and details subspace specifically helps to improve the speech intelligibility, despite there is no significant advantages over the complete spectrum.  


\subsection{Coherent and Incoherent Noise Reduction by the Proposed Algorithms}
In this section, different types of noise samples are applied to test the proposed TLSD-MS and SLSD-MS algorithms. Eight noise samples, including car, babble, construction crane, jet F16, fan noise, Volvo, machine gun, and white Gaussian, are used to represent various coherent ratio and SLR ( as shown in Fig.~\ref{snoise} and Table ~\ref{snoise1}).

\begin{figure*}[!hbt]
\vspace{-4mm}
\centerline{\includegraphics[scale=0.50]{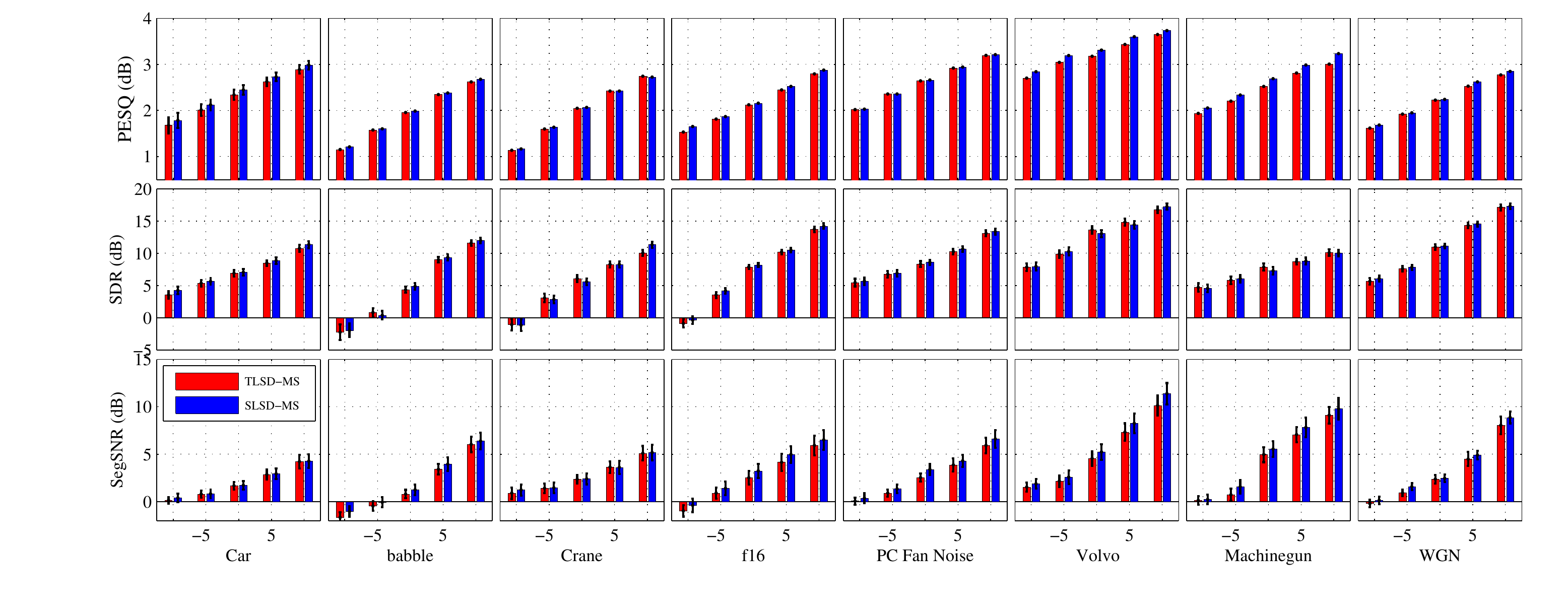}}
\vspace{-6mm}
\caption{Three metrics to evaluate the performances of the proposed TLSD-MS and SLSD-MS algorithms at different background noise scenario (8 different types of noise samples at various SNR levels (i.e., -10 ,-5 , 0, 5, 10 dB)).}
\vspace{-4mm}
\label{Two_comparison}
\end{figure*}

As shown in Fig.~\ref{Two_comparison}, the proposed algorithms achieve the best performance on Volvo noise, because this noise sample has the lowest coherence ratio (i.e., $C_{Volvo} = 0.01$). Generally, the performance of both algorithms decreases as the coherence ratio increases. For instance, the averaged performance of machinegun noise ($C_{machinegun} = 0.52$) is higher than that of babble noise ($C_{babble} = 0.85$), which is comparable with the results of crane noise ($C_{crane} = 0.81$). The reason is that high coherence ratio causes ambiguity on speech extraction. 

The results in Fig.~\ref{Two_comparison} also reveal that, not only coherent ratio can greatly affect the performance, the SLR also plays a role on noise cancellation in the proposed speech enhancement framework. It clearly shows that the averaged performance of jet F16 noise ($C_{F16} = 0.83$) is better than that of babble noise ($C_{babble} = 0.85$), despite the two noise samples have almost the same coherence ratios. The same situation happens for machinegun noise ($C_{machine gun} = 0.52$) and car noise ($C_{car} = 0.57$). The potential explanation may be that jet F16 and machinegun noise demonstrates either a superior low-rank ($R_{F16, Noise} = 0.92$) in the spectral envelop subspace or a more sparsity ($R_{Machinegun, Noise} = 0.01$,$R_{Machinegun, Activation} = 0.01$) in both the spectral envelop subspace and the speech dictionary space. These characteristics help the background noise more distinguishable, and by utilizing SLR, the proposed algorithms can impose low-rank and sparse constraints to separate the speech and noise components successfully. 

\subsection{Assessment via Speech Quality Metrics}
The performance of the proposed TLSD-MS and SLSD-MS algorithms are compared with four state-of-the-art speech enhancement algorithms, including MMSE-SPP \cite{hendriks2013dft}, NMF-RPCA \cite{chen2013speech}, RPCA \cite{huang2012singing}, and LARC \cite{sigg2012speech}. The evaluation is implemented across 25 noise samples at various SNRs (-10,-5, 0, 5, and 10 dB). Each benchmark algorithm is fine-tuned to be one of the best alternatives. 

\begin{figure}[!hbt]
\vspace{-4mm}
\centerline{\includegraphics[scale=0.55]{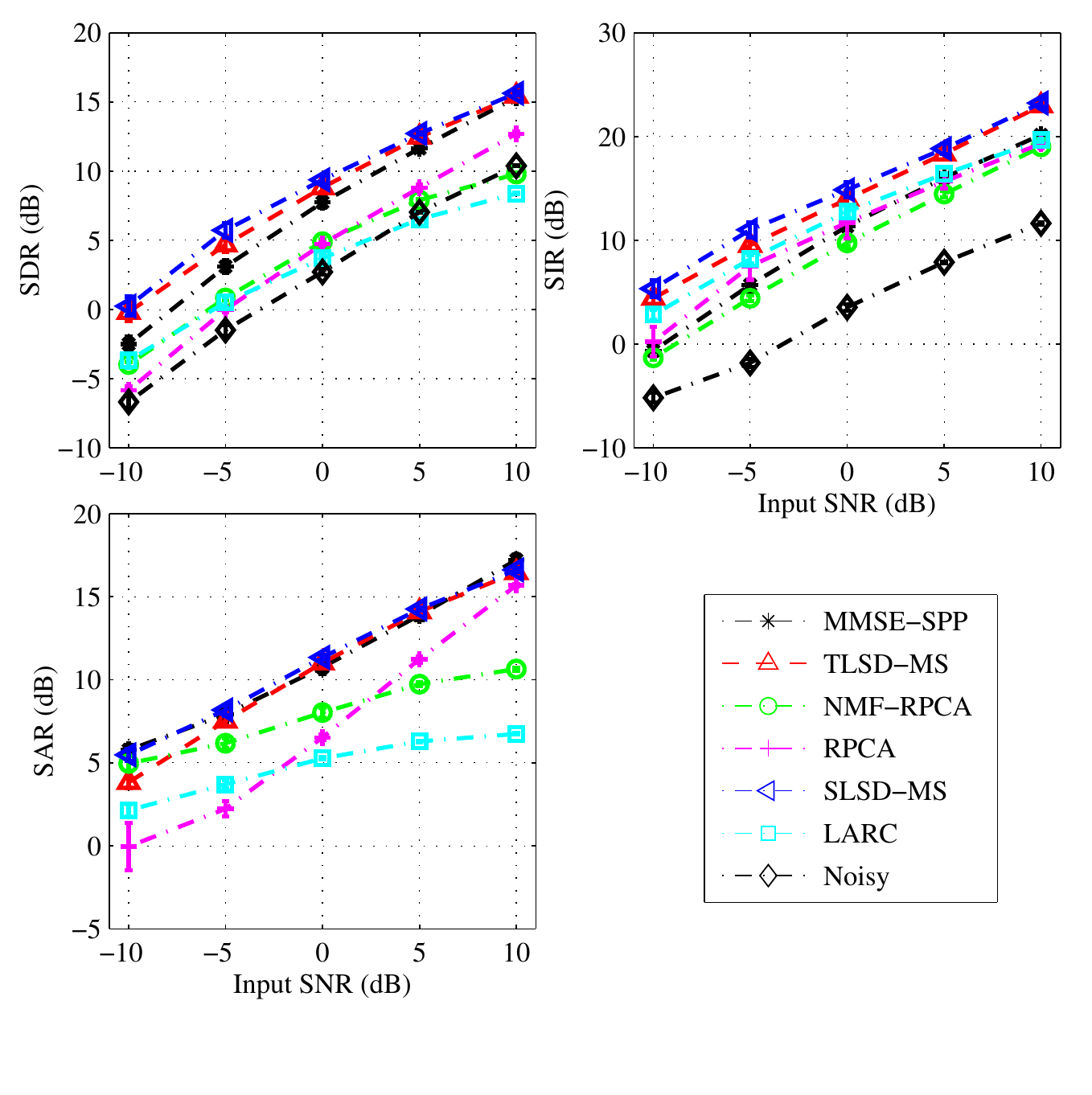}}
\vspace{-7mm}
\caption{SDR, SIR, and SAR of two proposed algorithms (TLSD-MS and SLSD-MS) and four other algorithms at various SNRs (-10, -5, 0, 5, and 10 dB). The results are averaged over 25 different types of noise samples.}
\vspace{-4mm}
\label{SSS}
\end{figure}

Figure~\ref{SSS} shows performance evaluation by three speech quality metrics, including the source to distortion ratio (SDR), source to interference ratio (SIR), and source to artifact ratio (SAR) from the BSS-Eval toolbox \cite{vincent2006performance}. SDR measures the overall quality of the enhanced speech, whereas SIR and SAR are proportional to the amount of noise reduction and inverse of the speech distortion. The results show that both proposed LSD-MS algorithms take advantages over four other algorithms with respect to all three metrics. In addition, SLSD-MS algorithm demonstrates slightly better performance than that of TLSD-MS algorithm. The underlying reason could be that a large part of selected noise samples have low SLR values, which are suitable for SLSD-MS algorithm. 

\begin{figure}[!hbt]
\vspace{-4mm}
\centerline{\includegraphics[scale=0.55]{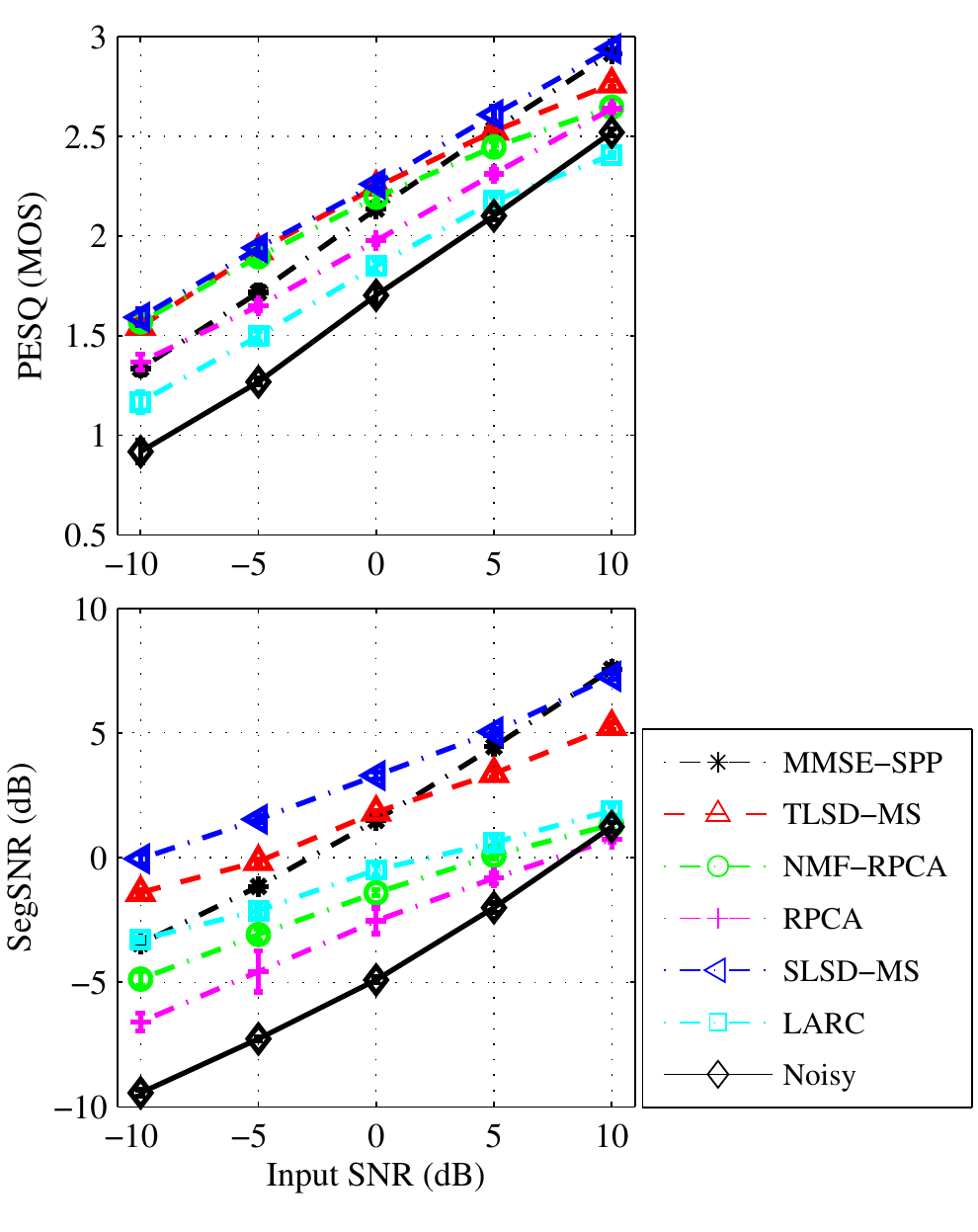}}
\vspace{-4mm}
\caption{PESQ and SegSNR of two proposed algorithms (TLSD-MS and SLSD-MS) and four other algorithms at various SNRs (-10, -5, 0, 5, and 10 dB). The results are averaged over 25 different types of noise samples.}
\vspace{-4mm}
\label{PSS}
\end{figure}

Evaluation results using SegSNRs and PESQ are shown in Fig.~\ref{PSS}. Both TLSD-MS and SLSD-MS algorithms outperform four other algorithms in terms of PESQ and SegSNR, and this superiority is especially significant when compared with unsupervised methods (i.e., MMSE-SPP and RPCA) at low SNRs. As a supervised technique, NMF-RPCA also achieves a good performance at low SNRs. The dictionary based speech recovery techniques can more effectively extract the speech components when speech is severely corrupted by background noise. Specifically, two proposed algorithms utilize the structure characteristics of speech and noise spectrogram in two modulated subspaces, and successfully avoid the general issues of dictionary based speech enhancement, such as overfitting and speech-like noise. 

\subsection{Assessment via Speech Intelligibility Metrics} 
To evaluate the performance of proposed algorithms on intelligibility of enhanced speech, three popular indexes, including hearing-aid speech quality index (HASQI) \cite{kates2014hearing}, normalized covariance metric (NCM) \cite{ma2009objective}, and short-time objective intelligibility (STOI) \cite{taal2011algorithm}, are employed in this section. HASQI has great potential to specifically capture quality when speech is subjected to a wide variety of distortions. This index can accurately predict the speech intelligibility ratings and generally as an improved version of Coherence speech intelligibility index (CSII). NCM is similar to the speech-transmission index (STI), and it computes the STI as a weighted sum of transmission index values determined from the envelops of the probe and response signals in each frequency band. STOI is also applied to validate the short time segmentation of the enhanced speech. All three metrics are expected to have a monotonic relation with the subjective speech-intelligibility, where a higher value denotes better intelligible speech.

\begin{figure}[!hbt]
\vspace{-2mm}
\centerline{\includegraphics[scale=0.55]{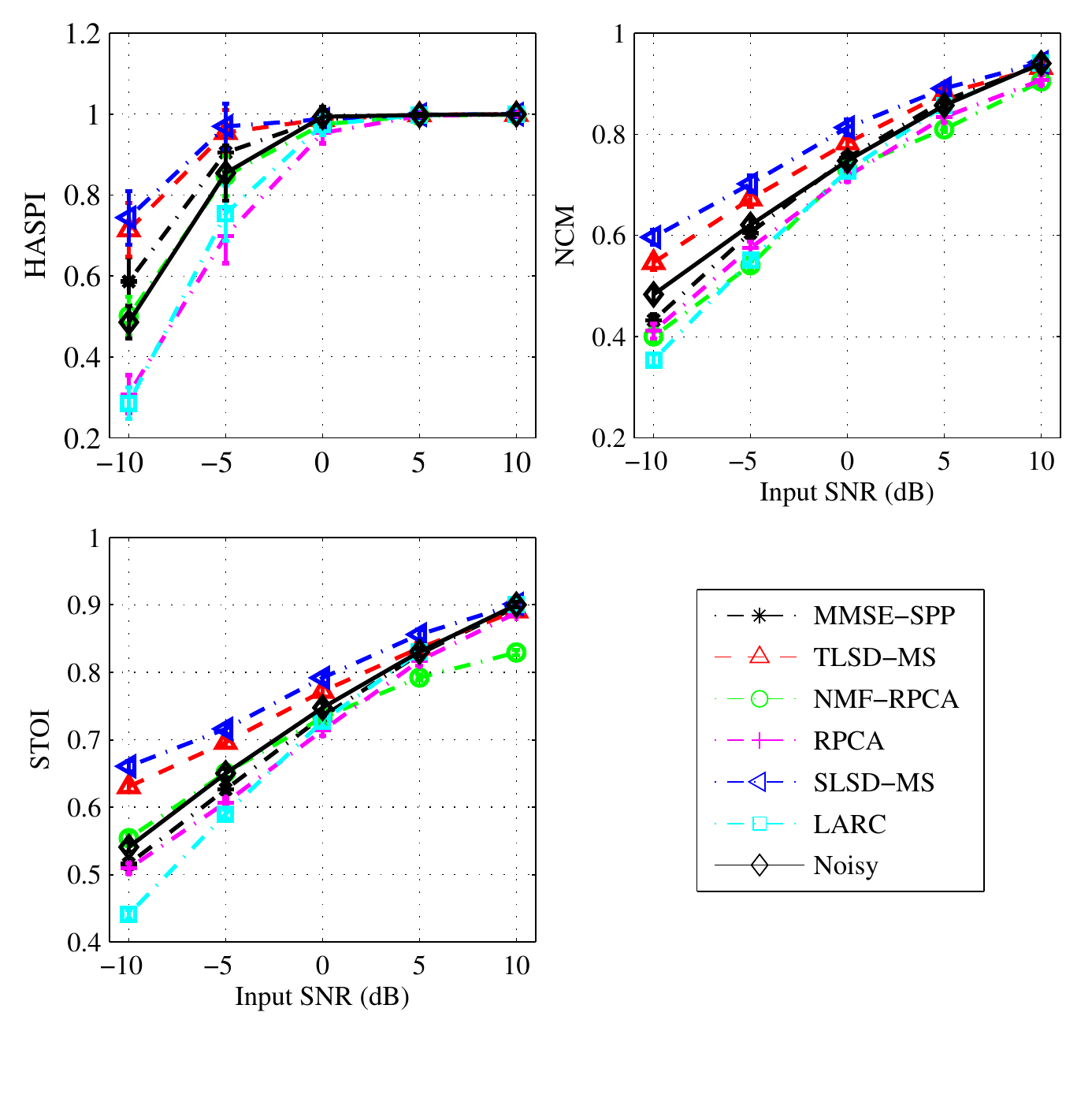}}
\vspace{-7mm}
\caption{HASPI, NCM, and STOI of two proposed algorithms (TLSD-MS and SLSD-MS) and four other algorithms at various SNRs (-10, -5, 0, 5, and 10 dB). The results are averaged over 25 different types of noise samples.}
\vspace{-4mm}
\label{NSH}
\end{figure}

Figure~\ref{NSH} shows the speech intelligibility evaluations of the proposed algorithms compared with four other state-of-the-art algorithms at various SNRs. Both proposed algorithms demonstrate superiority on all three intelligibility metrics. Specifically, at low SNRs (-10 and -5 dB), significant intelligibility improvements have been achieved by our proposed algorithms when compared with those benchmark algorithms. Specifically, for two other supervised algorithms, NMF-RPCA and LARC, their intelligibility improvements degrade greatly at low SNRs. One reason is that our proposed algorithms learn dictionaries from the spectral envelop subspace, which avoids the interference from the spectral details subspace. The benefit is that when the noise level increases, the energy in the spectral details subspace can produce a biased approximation to both speech and noise components. Another substantial explanation is that our supervised algorithms mainly focus on the recovery of the spectral envelop of speech, which is directly associated with speech intelligibility.  
In addition, the SLDS-MS algorithm demonstrates better intelligibility improvements than the TLSD-MS algorithm. Because, for noise with small R\(_{Activation}\), the tight low-rank constraints imposed by SLDS-MS works better to separate the low-rank noise components and sparse speech components in the spectral envelop subspace.

\section{Conclusion}
In this paper, we proposed a novel modulation subspace based speech enhancement framework. An acoustic model referred as formant-and-pitch was applied to obtain the spectral envelop and spectral details subspaces, in which supervised and unsupervised low-rank and sparse decompositions were implemented, respectively. To obtain the speech dictionary in the spectral envelop subspace, BNMF was utilized to inherently capture the temporal dependencies. Two different LDS schemes in the spectral envelop subspace were developed. By imposing different forms of norm to constraint rank and sparsity, the two approaches aimed to be adaptive to various background noise. The performance of the two developed algorithms were compared with other four existing speech enhancement algorithms, including MMSE-SPP \cite{hendriks2013dft}, NMF-RPCA \cite{chen2013speech}, RPCA \cite{huang2012singing} and LARC \cite{sigg2012speech}. Results showed that our developed algorithms not only showed robust performance under different background noise, but also achieved remarked improvements on speech perceptional quality with respect to various metrics. In addition, considerably robust performance is also demonstrated for different speech dictionaries obtained from several databases in spectral envelop subspace. Results showed that the MS based LDS approaches demonstrated significant improvements on speech intelligibility, when compared with other state-of-the-art algorithms.

\section{Acknowledgments}
This research was supported in part by the Illinois Clean Coal Institute (ICCI) with funds made available by the State of Illinois.

\bibliographystyle{IEEEtran}
\bibliography{mybib}


%

\end{document}